\documentclass[aps,prd,onecolumn,notitlepage,preprint,showpacs, showkeys,superscriptaddress]{revtex4-2}

\usepackage{longtable}
\usepackage{graphicx}
\usepackage{epstopdf}
\usepackage{amsmath}
\usepackage[section]{placeins}
\usepackage{slashed}
\usepackage{color}
\usepackage{soul}
\usepackage{appendix}
\usepackage{placeins}
\usepackage[utf8]{inputenc}
\usepackage{pstricks}
\usepackage{multirow}
\usepackage{hyperref}
\usepackage{subfigure}
\usepackage{epsfig}
\usepackage{cancel}
\usepackage{braket}
\usepackage[normalem]{ulem} 
\usepackage{longtable,booktabs}
\newcommand{\be}{\begin{equation}}
\newcommand{\ee}{\end{equation}}
\newcommand{\ben}{\begin{eqnarray}}
\newcommand{\een}{\end{eqnarray}}

\usepackage{ulem}
\usepackage{float}

\begin{document}

\title{Enigmatic properties of $\Xi(1620)$ and $\Xi(1690)$}

\author{Ta\'isa Veloso}
\email[]{taisa.veloso@unifesp.br}
\affiliation{Universidade Federal de S\~ao Paulo, C.P. 01302-907, S\~ao Paulo, Brazil}

\author{K.~P.~Khemchandani}
\email[]{kanchan.khemchandani@unifesp.br}
\affiliation{Universidade Federal de S\~ao Paulo, C.P. 01302-907, S\~ao Paulo, Brazil}

\author{A.~Mart\'inez~Torres}
\email[]{amartine@if.usp.br}
\affiliation{Instituto de F\'{\i}sica, Universidade de S\~{a}o Paulo, Rua do Mat\~{a}o, São Paulo SP, 05508-090, Brazil}   

\author{Luciano M. Abreu}
\email[]{luciano.abreu@ufba.br}
\affiliation{Instituto de Física, Universidade Federal da Bahia, Campus Ondina, Salvador, Bahia 40170-115, Brazil} 

\author{Li-Sheng Geng}
\email{lisheng.geng@buaa.edu.cn}
\affiliation{Sino-French Carbon Neutrality Research Center, \'Ecole Centrale de P\'ekin/School of General Engineering, Beihang University, Beijing 100191, China}
\affiliation{School of Physics, Beihang University, Beijing 102206, China}
\affiliation{Peng Huanwu Collaborative Center for Research and Education, Beihang University, Beijing 100191, China}
\affiliation{Southern Center for Nuclear-Science Theory (SCNT), Institute of Modern Physics, Chinese Academy of Sciences, Huizhou 516000, China}

\vspace{1cm}

\begin{abstract}
Motivated by a considerable release of data on first two excited states of the doubly strange baryon, $\Xi(1620)$ and $\Xi(1690)$, from different experimental collaborations in recent years, we investigate meson-baryon interactions in a coupled-channel approach.  The systems are composed of  pseudoscalar and vector mesons and the calculations are done for the total spin 1/2 and isospin 1/2 configuration. The tree-level interactions are deduced from Lagrangians based on chiral and hidden local symmetries. Specifically, we make an attempt to describe (1) data on the $K^-\Lambda$ femtoscopic correlation function coming from $p\,p$ and $Pb\,Pb$ collisions, and (2) the available data on invariant mass distributions for different systems: $\pi\Xi$, $\bar K \Sigma$ and $\bar K \Lambda$. We find that the amplitude which yields the correlation function  close to the data results in mass distributions of different meson-baryon systems that are in discordance with the experimental data and vice-versa. 
\end{abstract}

\maketitle
\newpage
\section{\label{sec:level1} Introduction}
In the present work, our aim is to investigate the controversies accumulating on the properties of the first excited baryons with strangeness $-2$: $\Xi(1620)$ and $\Xi(1690)$. We are particularly interested in calculating observables and comparing them with the data brought forward by  experimental studies that emerged in recent years and providing some clarification on the puzzles posed by those measurements. The current work is an extension of Ref.~\cite{Khemchandani:2016ftn}, which suggested that $\Xi(1690)$ can be understood as a state arising from the interplay of  pseudoscalar-baryon and vector-baryon dynamics.  We also follow Ref.~\cite{Ramos:2002xh}, where an important relation of hadron dynamics to understanding the properties of $\Xi(1620)$ was discussed.
 The results of the former works are very useful, given that the first excited state of $\Xi$  predicted by the quark model~\cite{Capstick:1986ter} has a mass around 1800 MeV, while two states with lower masses have been reported by different experimental collaborations~\cite{Borenstein:1972sb,Ross:1972bf,Briefel:1977bp,Hassall:1981fs,Biagi:1986zj,Belle:2001hyr,BESIII:2015dvj,Belle:2018lws,ALICE:2020wvi,LHCb:2020jpq,ALICE:2023wjz,BESIII:2023mlv}.  

 We would now like to summarize the findings of recent experimental investigations to show that the properties of light cascades as inferred from the different data seem to be incompatible with each other. Conclusions along the same line were drawn earlier in Ref.~\cite{Nishibuchi:2023acl} while focusing on data on $\Xi(1620)$. The idea here is to show that similar conflicts can be seen when considering  more data. We can begin by stating that $\Xi(1620)$ started receiving attention after the BELLE collaboration~\cite{Belle:2018lws}, in 2019, observed the neutrally charged $\Xi(1620)$ in the decay $\Xi^{+}_{c}\to \Xi^- \pi^+ \pi^+$ and reported the mass ($M$) and width ($\Gamma$) to be $1610.4 \pm 6 ^{+5.9}_{-3.5}$ MeV and $59.9 \pm 4.8^{+2.8}_{-3.0}$ MeV, respectively. Curiously, the mass value in Ref.~\cite{Belle:2018lws} is compatible with the one listed by the Particle Data Group (PDG)~\cite{PDG2024} but the width is larger than the one found in previous works~\cite{Ross:1972bf,Briefel:1977bp}, and larger than those of the other known $\Xi$ excited states. 

More recently, the ALICE collaboration~\cite{ALICE:2023wjz} announced the first observation of $\Xi(1620)^-$ decaying to $K^-\Lambda$ in $pp$-collisions using the femtoscopic technique. The collaboration determined the properties of the negatively charged $\Xi(1620)$ through a fit and obtained $M=1616.34^{+0.01}_{-0.05}$ MeV and $\Gamma=12.00\pm1.24$ MeV. Despite the compatibility with the masses reported by Refs.~\cite{Belle:2018lws,ALICE:2023wjz} when considering error bars, the values of the width are clearly different. 

To worsen the scenario, the BESIII collaboration has  obtained two sets of data on the invariant mass of $K^-\Lambda$ in $\psi(3686)\to K^-\Lambda \bar{\Xi}^++\text{c.c.}$\cite{BESIII:2015dvj,BESIII:2023mlv} and does  not find any signal of $\Xi(1620)$, even with four times larger statistics collected in the more recent experiment~\cite{BESIII:2023mlv}. This observation could imply that $\Xi(1620)$ lies below the $K^-\Lambda$ threshold. Now, in the same data sets, a clear evidence for $\Xi(1690)$ is seen, though with different values for its  width  determined in the two experiments (see Table~\ref{bes}).
\begin{table}[h!]
    \centering
    \caption{Properties of the $\Xi(1690)$ reported by the BESIII collaboration~\cite{BESIII:2015dvj,BESIII:2023mlv,BESIII:2026jcv}.}
    \begin{tabular}{c|p{3.5cm} p{3.5cm}}
     \hline \hline
      Year  & \, Mass (MeV) & Width (MeV)  \\ \hline
      2015~\cite{BESIII:2015dvj}  & $1687.7\pm 3.8 \pm 1.0$ & $27.1 \pm 10.0 \pm 2.7 $ \\
      2024~\cite{BESIII:2023mlv}  & $1685^{+3}_{-2} \pm 12.0$ & $81^{+10}_{-9} \pm 20 $ \\ 2026~\cite{BESIII:2026jcv}  & $1695.0\pm 9$ & $12.8 \pm 1.8  $ \\ \hline \hline
    \end{tabular}
    \label{bes}
\end{table}

Very recently, BESIII has also reported high statistics  data on $J/\psi\to K^-\Sigma^0\bar \Xi^+$~\cite{BESIII:2026jcv} and determined the mass and width of $\Xi(1690)$ as  $M=1695\pm 9$ MeV and $\Gamma=12.8\pm1.8$ MeV, respectively. This latest value of the width of $\Xi(1690)$ is in agreement with that given in the first row in Table~\ref{bes} (though not with the data cited in the second row).

Information on this sector has also been brought forward by the LHCb collaboration~\cite{LHCb:2020jpq}, through data on the $K^-\Lambda$ invariant mass spectra, obtained in the process $\Xi_b^-\to J/\psi \Lambda K^-$. By performing a full amplitude analysis, the collaboration extracted the mass and width of $\Xi(1690)^-$ to be $1692.0\pm1.3^{+1.2}_{-0.4}$ MeV and $25.9\pm9.5^{+14.0}_{-13.5}$ MeV, in agreement with the previous work of BESIII~\cite{BESIII:2015dvj} and with the more recent one~\cite{BESIII:2026jcv}, although not with the 2024 results~\cite{BESIII:2023mlv}. Interestingly, no evidence for $\Xi(1620)$ is found in Ref.~\cite{LHCb:2020jpq}, as also pointed out by the authors of Ref.~\cite{Feijoo:2024qqg}.

In the present work, we confront almost all the data determined in the last two decades with our model and find that there are two sets of conflicting data, and only one of them can be satisfactorily described at the same time. In the next section, we discuss the details of the model and, subsequently, show and interpret the results obtained in this work.

\section{Theoretical framework}
In this section, we aim to summarize the details of the model used in our work. First, we present a discussion on the evaluation of scattering amplitudes obtained by solving the Bethe-Salpeter equation. Next, we describe how the calculation of the femtoscopic correlation function of $\bar K \Lambda$ has been carried out using the model amplitudes, where we follow the formalism of the ALICE Collaboration~\cite{ALICE:2020wvi,ALICE:2023wjz} in order to make a comparison with the experimental data determined in the mentioned works.

\subsection{Deducing scattering amplitudes}
The basic idea behind the model is to consider interactions among different meson-baryon systems that couple to strangeness $-2$, determine kernel amplitudes, solve the scattering equation in a coupled-channel approach, and use the resulting amplitudes to calculate observables that can be compared with the available experimental data. The coupled-channel basis in our model consists of $\pi \Xi$, $\eta \Xi$, $\bar K \Sigma$, $\bar K \Lambda$, $\rho\Xi$, $\omega\Xi$, $\phi\Xi$, $\bar K^* \Sigma$, and $\bar K^* \Lambda$. 

\begin{figure}[h!]
    \centering
    \includegraphics[width=0.78\linewidth]{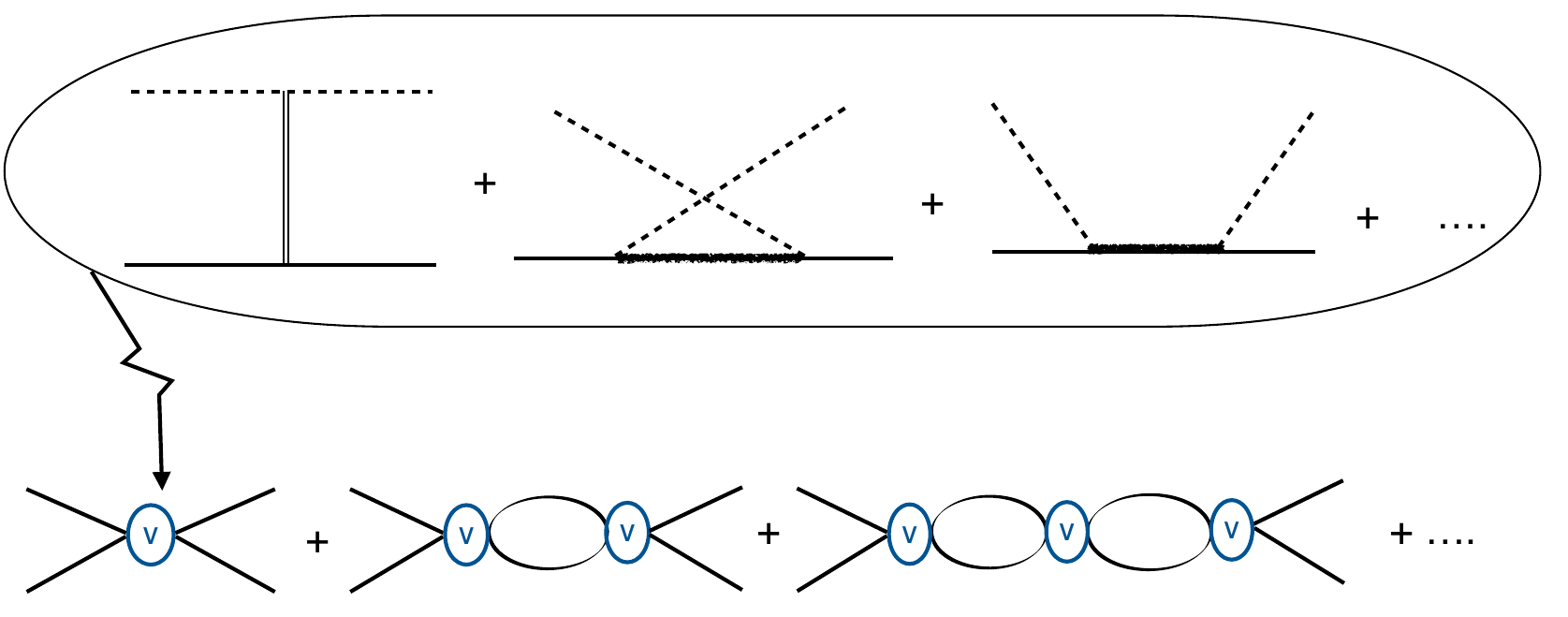}
    \caption{Feynman diagrams that  contribute to the meson-baryon scattering at the lowest order are shown in the upper row. The fuzzy thick line represents the exchange of a baryon (a cascade in the $s$-channel, and a $\Lambda/\Sigma/\Xi/N$ in the $u$-channel). The sum of the diagrams is used as a kernel to solve the scattering equation depicted schematically in the lower row.}
    \label{diagram}
\end{figure}

The scattering amplitude for each $i\to j$ process is determined by calculating different diagrams contributing to it, such as an $s$- and $u$-channel exchange of an octet baryon, a $t$-channel exchange of a vector meson, and a contact term, as shown in Fig.~\ref{diagram}. The sum of such diagrams is treated as a kernel to solve the Bethe-Salpeter equation (see Fig.~\ref{diagram}) 
\begin{equation}\label{bseq}
    T_{ij}=V_{ij}+ \sum_l V_{il}\,G_{l}\,T_{lj}
\end{equation}
in a coupled-channel approach.

Our formalism is based on Lagrangians formulated in terms of hadronic effective degrees of freedom, while preserving the relevant symmetries of QCD. The well-established chiral Lagrangian~\cite{Gasser:1983yg} provides an appropriate effective field theory to describe interactions in systems composed of pseudoscalar mesons. We rely on the lowest-order chiral Lagrangian~\cite{Oset:1997it,Oller:2000fj}, the relevant part of which can be written as
\begin{align}
    \mathcal{L}_{PB}=\frac{i}{8f^2} \bar B \gamma_\mu\left[\left[P,~\partial_\mu P\right],B\right]-\frac{1}{2f}\Biggl(D^\prime\langle\bar B\gamma_\mu\gamma_5\left\{\partial_\mu P,~B\right\}\rangle+F^\prime\langle\bar B\gamma_\mu\gamma_5\left[\partial_\mu P,~B\right]\rangle\Biggr),\label{PBlag}
\end{align}
where $P$ and $B$ stand for the SU(3) matrices of the pseudoscalar meson and octet baryon fields, respectively, and $D^\prime=0.8$, $F^\prime=0.46$~\cite{Khemchandani:2011mf}. The first term of $ \mathcal{L}_{PB}$ gives rise to a contact interaction, while the second part describes the Yukawa vertices that contribute to the $s$- and $u$-channel diagrams. 

The amplitudes related to the contact interaction were first determined in Ref.~\cite{Ramos:2002xh}, and we have obtained the same expressions for the different pseudoscalar-baryon coupled channels. Hence, we do not state them here. 

We obtain the expressions for the $s$- and $u$-channel exchange of octet-baryons, using Eq.~(\ref{PBlag}), and after projecting them on $S$-wave, we get
\begin{align}\label{sijeq}
    V_{ij,\,\text{PB}}^{s}=\frac{N_iN_j}{4f_i\,f_j}\frac{S_{ij}}{\sqrt{s}+M_{\Xi}}(\sqrt{s}-M_i)(\sqrt{s}-M_j),
\end{align}
and 
\begin{align}\label{uijeq}
    V_{ij,\,\text{PB}}^{u}&=-\frac{N_iN_j}{4f_i\,f_j}\sum_k \frac{U^k_{ij}}{u-M_k^2}\Big( u\big[\sqrt{s}+M_k\big] +\sqrt{s}\big[M_j( M_i+M_k )\big] + M_iM_k \nonumber\\
    &\quad-M_j\big[ M_i+M_k \big]\big[ M_i+M_j \big] -M_i^2M_k\Big),
\end{align}
where $S_{ij}$ and $U^k_{ij}$ are $SU(3)$ coefficients whose values are listed in Tables~\ref{sijk} and \ref{uijk}, respectively, for different processes. Here, and throughout the manuscript, the variables $M_i$ and $M_j$ represent the masses of the incoming and the outgoing baryons, and $f_k$ corresponds to the decay constant of the meson in the $k$th channel (see Table~\ref{decay}). \begin{table}[h!]
    \centering
    \caption{Decay constants for each meson as used in our scattering amplitudes calculation.}
    \begin{tabular}{c|c | c | c}
    \hline \hline
     Pseudoscalar & Decay constant (MeV) & Vector & Decay constant (MeV)\\ \hline
     $\pi$ & $93$ & $\rho$ & $153.45$ \\ 
     $\eta$ & $123.69$ & $\omega$ & $153.45$ \\
     $\bar K$ & $113.46$ & $\phi$ & $168.33$ \\
     $~$ & $~$ & $\bar K^*$ & $159.96$ \\ \hline \hline
    \end{tabular}
    \label{decay}
\end{table}
Clearly, only strangeness $-2$ baryons can be exchanged in the $s$-channel,  while a nucleon, a $\Xi$, a $\Lambda$ or a $\Sigma$ can be exchanged in the $u$-channel. The summation in Eq.~(\ref{uijeq})  accounts for the exchange of all possible octet baryons contributing to each process. Another pertinent remark to be made at this point is that the expression in Eq.~(\ref{uijeq}) corresponds to considering the center of mass momentum to vanish, which is consistent with a nonrelativistic limit considered in this work. Such an approximation is made mainly to avoid the left-hand cut. It is important to mention that we do not make this approximation for $\pi \Xi$ elastic scattering and follow Refs.~\cite{Khemchandani:2018amu,Veloso:2026kii} for this case. 

In Eqs.~(\ref{sijeq}) and (\ref{uijeq}), and in all the  subsequent expressions in this manuscript,  $N_i(N_j)$ stands for the normalization of the Dirac spinor which, in the convention we follow, is
\begin{align}
    N_l=\,\sqrt{\frac{E_l+M_l}{2\,M_l}}.
\end{align}

Next, we deduce the vector-baryon amplitudes using the Lagrangian~\cite{Khemchandani:2011et,Khemchandani:2011mf}
\begin{align}
    \mathcal{L}_{VB}&=-g\Biggl(\langle\bar B\gamma_\mu V^\mu B-B\gamma_\mu B \,V^\mu\rangle+\langle\bar B\gamma_\mu B\rangle\langle V^\mu\rangle \Biggr.\\
   &\quad \left.+\frac{D+F}{4M}\langle \bar B \sigma_{\mu\nu}V^{\mu\nu}B\rangle+\frac{D-F}{4M}\langle \bar B \sigma_{\mu\nu}BV^{\mu\nu}\rangle\right),
\end{align}
where $V^{\mu\nu}=\partial^\mu V^\nu-\partial^\nu V^\mu+ig\left[V^\mu,V^\nu\right]$, $D=2.4$, $F=0.82$, $\gamma_\mu$ are the Dirac matrices, $\sigma_{\mu\nu}=i[\gamma_\mu,\gamma_\nu]/2$ (see Refs.~\cite{Khemchandani:2011et,Khemchandani:2011mf,Khemchandani:2016ftn} for more details). A detailed study of the strangeness $-2$ sector was presented in Ref.~\cite{Khemchandani:2016ftn}, where Bethe-Salpeter equations were solved with fully relativistic kernels determined for vector-baryon channels. It was shown in the former work~\cite{Khemchandani:2016ftn} that the nonrelativistic treatment led to the same conclusions. Hence, we employ nonrelativistic amplitudes for the different diagrams contributing to vector-baryon interactions in this work. We find it instructive to provide a summary of all contributions here, since following the relations between relativistic and nonrelativistic amplitudes given in Ref.~\cite{Khemchandani:2016ftn} can be tedious for the reader. 

The $t$-channel amplitude in the $L=0$ partial wave, as first determined in Ref.~\cite{Oset:2010tof}, has the form
\begin{align}
    V^t_{ij, VB}=-\mathcal{A}_{ij}\frac{N_i N_j}{4 f_i f_j}\left(2\sqrt{s}-M_i -M_j\right)\vec \epsilon_i\cdot\vec \epsilon_j,\label{VB:t}
\end{align}
where $\epsilon_k$ represents the polarization of the vector meson in the $k$th channel and the values of $\mathcal{A}_{ij}$ for different processes are given in Table~\ref{aij} in the Appendix (which coincide with the values in Refs.~\cite{Oset:2010tof,Khemchandani:2016ftn}). The spin structure of this amplitude gives an identical contribution to the spin 1/2 as well as spin 3/2 cases. 

The contact interaction arising from the last term of $V^{\mu\nu}$, in the $s$-wave, leads to
\begin{align}
    V_{ij,VB}^{CT}=i \mathcal{B}_{ij}N_i N_j\frac{g_ig_j}{2\sqrt{M_i M_j}}~\vec\sigma\cdot(\vec \epsilon_2\times\vec \epsilon_1),\label{VB:CT}
\end{align}
where the coupling $g_k=m_k/(\sqrt{2}f_k)$ is defined in terms of the mass of the meson in the $k$th channel and the values of $\mathcal{B}_{ij}$ are given in Table~\ref{bij} in the Appendix.

The amplitudes for the $s$- and $u$-channel exchange of an octet baryon have a generic form
\begin{align}\label{VB:su}
    V_{ij,VB}^s&=N_iN_j~\mathcal{C}_{ij}\left(\frac{g_i g_j}{2\tilde M+\tilde m}\right)\vec \epsilon_j\cdot\vec \sigma~\vec \epsilon_i\cdot\vec \sigma,\\[2mm]\nonumber
    V_{ij,VB}^u&=N_iN_j~\mathcal{D}_{ij}\left(\frac{g_i g_j}{2\tilde M-\tilde m}\right)\vec \epsilon_i\cdot\vec \sigma~\vec \epsilon_j\cdot\vec \sigma,
\end{align}
with $\tilde M$ and $\tilde m$ being the SU(3) average masses of the baryons and mesons involved in the process. The values of $\mathcal{C}_{ij}$ and  $\mathcal{D}_{ij}$ are provided in Tables~\ref{cij} and \ref{dij} in the Appendix.

Finally, it remains to write the amplitude for the transition among  pseudoscalar-baryon and vector-baryon channels, which is 
\begin{align}\label{krolleq}
    V_{ij}^{KR}=iN_iN_j\frac{\mathcal{F}_{ij}}{2}\frac{g_{KR}}{\sqrt{f_if_j}},
\end{align}
where $g_{KR}$ is the Kroll-Rudermann coupling, defined as $g_{KR}=m_V/(\sqrt {2 \,f_Pf_V})$ following Ref.~\cite{Khemchandani:2016ftn}. The values of decay constants for each channel are given in Table~\ref{decay}. In the current work, we allow it to vary in a reasonable range in the $\chi^2$-fitting discussed in section~\ref{results}. The coefficients $\mathcal{F}_{ij}$ are provided in Table~\ref{fij} in the Appendix.

The last component needed to evaluate Eq.~(\ref{bseq}) is the loop function $G$, which is an integral of a meson and a baryon propagator (see Refs.~\cite{Oset:1997it,Oset:2010tof,Ramos:2002xh} for more details)
\begin{equation}
    G_j(\sqrt s)= \displaystyle\int \dfrac{d^4q}{(2\pi)^4}\frac{2M_j}{q^2 - m_j^2 +i \epsilon}\frac{i}{(p-q)^2-M_j^2+i\epsilon},
\end{equation}
with $p$ being the total four-momentum of the system.
Because of the divergent nature of the loop function, it is necessary to regularize it. We perform our calculations using a dimensional regularization method and write
\begin{align}\label{loop}
    G_j(\sqrt s)&=\frac{2M_j}{16\pi^2}\Bigg[a_j(\mu) + \ln \frac{m^2_j}{M^2_j}\,\frac{s-(M^2_j-m^2_j)}{2\,s}+\ln\frac{M^2_j}{\mu^2}  \\ \nonumber& + \frac{p_{cm}}{\sqrt s}\text{ln}\Bigg( \frac{(s-(M^2_j-m^2_j)+2\,p_{cm}\sqrt s\,)\,(s+(M^2_j-m^2_j)+2\,p_{cm}\sqrt s\,)}{(-s-(M^2_j-m^2_j)+2\,p_{cm}\sqrt s\,)\,(-s+(M^2_j-m^2_j)+2\,p_{cm}\sqrt s\,)} \Bigg) \Bigg],
\end{align}
which depends on the subtraction constants $a_i(\mu)$ at a given regularization scale $\mu$. Here $\mu$ is chosen to be $630 \text{ MeV}$, as in Refs.~\cite{Oset:1997it,Oset:2010tof,Ramos:2002xh}, and $a_i$ are considered as free parameteres in our model. It is important to mention that for channels composed of mesons with a finite decay width, in this case $\rho$ and $\bar K^*$, we must convolute this loop function by allowing the mass of the meson to vary within a certain interval (see Refs.~\cite{Oset:2010tof,Khemchandani:2016ftn,Khemchandani:2018amu} for more details).

\subsection{Determining the $\bar K \Lambda$ correlation function to be compared with the data from $p\,p$ and $Pb\,Pb$ collisions}

The femtoscopic correlation function (CF) has been gaining a lot of attention recently as a powerful new tool to investigate properties of non-conventional hadrons. Experimentally, it is defined as 
    \begin{equation}
        C(p)=\mathcal{K}\frac{N_\text{same}(p)}{N_\text{mixed}(p)},
    \end{equation}
where $\mathcal{K}$ is a normalization factor, $p$ is the relative momentum of a pair of hadrons under investigation, $N_\text{same}$ is the distribution of the  pair with the relative momentum $p$ detected in the same event, and $N_\text{mixed}$ is constructed by pairing particles from different events having the same relative momentum~\cite{ALICE:2020wvi,ALICE:2022yyh,ALICE:2023wjz}. 

It is important to recall that the measured CF does not result from the genuine contribution alone, and therefore one should also consider the existence of a residual contribution coming from other sources, like electromagnetic or weak decays, when comparing model results with the experimental data. For example, as mentioned in Ref.~\cite{ALICE:2020wvi}, the primary contribution is also made of particles coming from decay before the kinetic freeze-out. In the same work, the ALICE Collaboration performed a simulation and concluded that many different ``parent" pairs produce $\Lambda$-$\bar K$, which gives rise to a non-negligible ``residual" contribution.

Since our model focuses on the strong interaction between hadrons, those residual contributions should also be added to the final CF to compare the results with the data. In order to do so, we follow the relation given by the ALICE Collaboration~\cite{ALICE:2020wvi,ALICE:2023wjz}
    \begin{equation}
        C_\text{total}^{K^-\Lambda}\,(p)=\mathcal{N} \, C_\text{measured}(p) \times B(p),\label{exp_def}
    \end{equation}
where $\mathcal{N}$ is a normalization (which is $\sim$1) and $B$ is a non-femtoscopic background that can still give a contribution to the final experimental result and can vary with momentum, $p$. Because we intend to evaluate the $K^-\Lambda$ CF measured in two different types of collisions, (1) $Pb\,Pb$ and (2) $p\,p$, in order to make a comparison with the data in Refs.~\cite{ALICE:2020wvi,ALICE:2023wjz}, we must consider a different background in each calculation. In  case (2), we take almost the same background as given by the collaboration~\cite{ALICE:2023wjz}, except for the region where a signal related to $\Xi(1690)$ is present, since we expect this state to be generated from hadron-hadron interactions~\cite{Khemchandani:2016ftn,Sarti:2023wlg}. Meanwhile, for the case (1), we use the same background as given by the collaboration~\cite{ALICE:2020wvi}, where  $\Xi(1690)$ is not added.

Further, the measured CF is written as~\cite{ALICE:2020wvi,ALICE:2022yyh,ALICE:2023wjz} 
    \begin{align}
        C_\text{measured}\,(p)=1+\sum_{\alpha} \lambda_\alpha\big(C_{\alpha}\,(p)-1\big),
    \end{align}
where $\alpha$ denotes the genuine or the residual contribution. As mentioned earlier in this manuscript, in Ref.~\cite{ALICE:2020wvi}, the authors performed a simulation to estimate the values of the parameters $\lambda_\alpha$, which correspond to the proportion of pairs of $K^-\Lambda$ coming from the source $\alpha$.  It is important to mention that the experimentalists assumed the residual contribution to be flat.

In this manner, we can write
\begin{equation}
    C_\text{measured}^{K^-\Lambda}\,(p)=1+0.51(C_\text{gen}^{K^-\Lambda}\,(p_i)-1),
\end{equation}
using $\lambda_\text{gen}=0.51$ as determined in Refs.~\cite{ALICE:2020wvi,ALICE:2023wjz}.

At this point, we can start our discussion on how to evaluate $C_\text{gen}^{K^-\Lambda}$ from a theoretical point of view.  We use the Koonin-Pratt (KP) formula adapted for a coupled-channel approach~\cite{Vidana:2023olz} (some recent examples on the application of this expression can be found in Refs.~\cite{Feijoo:2023sfe,Albaladejo:2023pzq,Khemchandani:2023xup,Liu:2024uxn,Abreu:2024qqo,Liu:2025wwx,Veloso:2026kii}), 
\begin{align}\label{cgen}\nonumber
    C_\text{gen}^{K^-\Lambda}\,(p)= \int d^3r\, \mathcal{S}(r,R)|\Psi(\vec{p}\,,\vec{r})|^2  
    =1+4\pi \Theta(q_{max}-p)\int_0^\infty dr \, r^2  \mathcal{S}(r,R) \\[2mm] 
    \times \Big\{ \sum_j\omega_j\, |j_0(p\,r)\delta_{\left(K^-\Lambda\right)\,j}+T_{j\to K^-\Lambda}(\sqrt{s})\tilde{G}_j(r, \sqrt{s},\,q_{max})|^2 -j_0^2(p\,r)\Big\},
\end{align}
where $j$ represents a particular channel,  $j_0$ is the spherical Bessel function, and $T_{j\to K^-\Lambda}$ is the scattering amplitude of the $j$th channel producing $K^-\Lambda$, evaluated by solving Eq.~\ref{bseq}. Next, we will discuss the description of the source function, $\mathcal{S}(r,R)$, and the determination of the weights, $\omega_j$, for different channels. Before going to those discussions, it is important to mention that the loop function $\tilde{G}_j(r,\sqrt{s},\,q_{max})$ is calculated as
\begin{equation}
    \tilde{G}_j(r,\sqrt{s},\,q_{max})= 2M_j\int\limits_0^{q_{max}}\frac{d^3q}{(2\pi)^3}\frac{E_j+w_j}{2E_j\,w_j}\frac{j_0(qr)}{s-(E_j+w_j)^2 + i\epsilon},
\end{equation}
where $E_j$ ($w_j$) stands for the energy of the baryon (meson) in the $j$th channel.

Let us recall that the source function depends on the collision. In the present case, where we are discussing the $K^-\Lambda$ CF coming from two different collisions ($p\,p$ and $Pb\,Pb$), we must carry out the calculations within two distinct descriptions for $\mathcal{S}(r,R)$, following Refs.~\cite{ALICE:2020wvi,ALICE:2023wjz},  as described below. 
\begin{enumerate}
    \item Source function for the $Pb\,Pb$ collision: In this case, no particular information on the profile of the source function is provided by the collaboration. Thus, we choose a Gaussian distribution, with $R=5\text{ fm}$,
    \begin{equation}
        \mathcal{S}(r,R)=\frac{1}{(4\pi)^{3/2} R^3} \text{e}^{-\frac{r^2}{4R^2}}.
    \end{equation}
    \item Source function for the $p\,p$ collision: In contrast to the previous case, the source function profile for the CF of charged kaons and $\Lambda$ in the $p\,p$ collision was specified in Ref.~\cite{ALICE:2023wjz}. Therefore, we follow the experimental collaboration and define the source as a weighted sum of two Gaussians that depends on $R_1 =1.202 \text{ fm}$ and $R_2 =2.330 \text{ fm}$~\cite{ALICE:2023wjz},
    \begin{equation}
        \mathcal{S}(r,R_1,R_2)=\lambda_{\mathcal{S}}\big[  \omega_{\mathcal{S}} \,\mathcal{S}_1(r,R_1)+(1-\omega_{\mathcal{S}})\,\mathcal{S}_2(r,R_2)\big],
    \end{equation}
    where $\lambda_{\mathcal{S}}=0.9806$ and $\omega_{\mathcal{S}}=0.7993$~\cite{ALICE:2023wjz}.
\end{enumerate}

The last  ingredients remaining to evaluate Eq.~(\ref{cgen}) are the production weights, $\omega_j$, for each coupled channel,
\begin{equation}\label{weights2}
    \omega_j=\frac{N_j^B N_j^M}{N_{K^-\Lambda}},
\end{equation}
where $N_j^B$ ($N_j^M$) is the primary production yield of the baryon (meson) in the $j$th channel~\cite{Sarti:2023wlg}, and $N_{K^-\Lambda}$ is the primary production yield of $K^-\Lambda$.

We make use of the statistical fireball model, following Ref.~\cite{Barbat:2025orm}, to evaluate the primary production yield for cases (1) and (2),
\begin{equation}\label{yields}
    \frac{dN}{m_{\bot}dm_{\bot}\,dy\, d\phi} \sim m_{\bot}\text{cosh}(y)\,\text{exp}\Bigg({-\frac{m_{\bot} \text{cosh}(y)}{T_{ch}}}\Bigg).
\end{equation}
Here, $T_{ch}$ is the chemical freeze-out temperature that varies with the centrality and the nature of the colliding pair. We take the values given in Ref.~\cite{ALICE:2022yyh} for the $p\,p$ collision studied at $\sqrt{s}=13$~TeV and for the lowest centrality for the $Pb\,Pb$ collision. Further, $y$, in Eq.~(\ref{yields}), is the rapidity, $m_{\bot}$ is the transverse mass of the particle for which the  production yield is evaluated and $\phi$ is the azimuthal angle associated to the momentum of the particle~\cite{Barbat:2025orm}. The values of $\omega_j$ are provided in Table~\ref{weights}. 
\begin{table}[h!]
    \centering
    \caption{Production weights obtained via the fireball model.}   \label{weights}
    \begin{tabular}{c|c c}
    \hline \hline
     Production weights & $p\,p$ collision & $Pb\,Pb$ collision \\
     \hline 
     $\omega_{\pi^-\Xi^0}$ & $1.12$ & $1.14$ \\
     $\omega_{\pi^0\Xi^-}$ & $1.09$ & $1.11$ \\
     $\omega_{\eta\Xi^-}$ & $0.30$ & $0.28$ \\
     $\omega_{K^- \Sigma^0}$ & $0.69$ & $0.68$ \\
     $\omega_{\bar K^0 \Sigma^-}$ & $0.68$ & $0.65$ \\
     $\omega_{K^- \Lambda}$ & $1$ & $1$ \\
     $\omega_{\rho^-\Xi^0}$ & $0.12$ & $0.11$ \\
      $\omega_{\rho^0\Xi^-}$ & $0.12$ & $0.11$ \\
     $\omega_{\omega\Xi^-}$ & $0.11$ & $0.10$ \\
     $\omega_{\phi\Xi^-}$ & $0.04$ & $0.04$ \\
     $\omega_{ K^{*-} \Sigma^0}$ & $0.13$ & $0.12$ \\
     $\omega_{\bar K^{*0} \Sigma^-}$ & $0.13$ & $0.11$ \\
     $\omega_{K^{*-} \Lambda}$ & $0.19$ & $0.17$ \\
     \hline\hline
    \end{tabular}
\end{table}

The results of the CF obtained for both cases,  using the scattering amplitudes given by our model, will be presented in the next section of the manuscript.

\section{Results}\label{results}
We begin by discussing the different data sets that have been determined in recent decades and which should be consistently described within the same model. 
\begin{itemize}
    \item Data on the invariant mass of $\pi^+\Xi^-$ are available from the BELLE Collaboration~\cite{Belle:2018lws}, which shows a signal for $\Xi(1620)$ and $\Xi(1690)$. It is relevant to mention that the signal significance for $\Xi(1620)$ has been found to be 25$\sigma$, while the evidence for $\Xi(1690)$ is limited to 4-4.5~$\sigma$ only.

    \item Several facilities have reported data on the $\bar K\Lambda$ invariant mass, as determined from different processes~\cite{Belle:2001hyr,BESIII:2015dvj,LHCb:2020jpq,ALICE:2023wjz,BESIII:2023mlv}. Interestingly, as shown in an explicit plot in Ref.~\cite{Feijoo:2024qqg}, data from  Refs.~\cite{LHCb:2020jpq,ALICE:2023wjz} seem to disagree in the energy region below 1670 MeV. We will plot (together with our results),  later in this section, data from LHCb~\cite{LHCb:2020jpq}, the recent work of BESIII~\cite{BESIII:2023mlv} and BELLE~\cite{Belle:2001hyr}, to show that those data sets are compatible with each other.

    \item The invariant mass spectra of $K^-\Sigma^+$ determined by the BELLE Collaboration in 2002~\cite{Belle:2001hyr}. It  is useful to mention here that a signal for $\Xi(1690)$ is seen in these data.

    \item Data on the femtoscopic correlation function of $K^-\Lambda$ determined in $p\,p$ collisions at $\sqrt{s}=13$ TeV~\cite{ALICE:2023wjz}.
    \item Data on the femtoscopic correlation function of $K^-\Lambda$ extracted from $Pb\,Pb$ collisions at $\sqrt{s_{NN}}=2.56$ TeV~\cite{ALICE:2020wvi} for various centralities. We will consider the data available for the centrality interval of $30\% - 50\%$ when comparing with the model results. This particular choice is motivated by the information available in Ref.~\cite{ALICE:2022yyh} on the temperature of the chemical freeze-out, $T_{ch}$, for $Pb\,Pb$ collisions, that is needed to calculate weights through Eq.~(\ref{weights2}).
    It is useful to add that the closest centrality interval  for which the information on $T_{ch}$ given in Ref.~\cite{ALICE:2022yyh} is for the range $60\% - 70\%$.
\end{itemize}

We find it useful to summarize our findings before going into the details. We started by making an attempt to fit almost all the aforementioned data sets simultaneously (we considered data on the invariant masses and CF from either $p\,p$ or $Pb\,Pb$ collisions). The parameters which we treat as free in the $\chi^2$-fit are (1) the subtraction constants needed to regularize the divergent loop for each of the 9 channels, (2) the three-momentum cut-off~($\Lambda$) present in the form factor in the kernel for the $s$- and $u$-channel diagrams [Eqs.~(\ref{sijeq}) and (\ref{uijeq})] to take into account the suppression of baryon-antibaryon annihilation vertices, (3) the cut-off ($q_{max}$) required to calculate the correlation function  in Eq.~(\ref{cgen}) (4)  the Kroll-Ruddermannn coupling, $g_{KR}$, that relates the transition among pseudoscalar-baryon and vector-baryon channels. 

We did not succeed in achieving an agreement with the entire set. We find that the lowest chi-square values always led to poor agreement with the data on the invariant mass of different systems and even a poorer description of the data on the correlation function. We then decided to fit data on the invariant mass spectra and the CF separately and ended up discovering that only one of the two can be fitted satisfactorily. 

To demonstrate our results clearly, we would like to show three different fit results where the data sets considered are summarized in Table~\ref{data}, labeled as ``Fit 1", ``Fit 2" and ``Fit 3".
\begin{table}[h!]
    \centering
    \caption{Data used for Fit 1, Fit 2 and Fit 3.}
    \begin{tabular}{c | c  c c}
    \hline \hline
    Data &$\quad$ Fit 1 $\quad$ & $\quad$Fit 2 $\quad$ & $\quad$ Fit 3 $\quad$ \\ \hline
    $\pi^+\Xi^-$ invariant Mass~\cite{Belle:2018lws} & X  &  X & \\
    $\bar K^0 \Lambda$ invariant Mass~\cite{Belle:2001hyr} & X & X & \\
    $ K^- \Sigma^+$ invariant Mass~\cite{Belle:2001hyr} & X &  X & \\
    $K^-\Lambda$ correlation Function in $p\,p$ collision~\cite{ALICE:2023wjz} & & & X \\
     $K^-\Lambda$ correlation Function in $Pb\,Pb$ collision~\cite{ALICE:2020wvi} &X & & 
     \\ \hline \hline     
    \end{tabular}
    \label{data}
\end{table}

Before showing the results, we must discuss how the different invariant masses have been calculated. In particular, we choose to use the data on the invariant mass of $K^-\Lambda$ provided by the BELLE Collaboration~\cite{Belle:2001hyr}, since more  data points are available in this set. Data have been determined for the process $\Lambda^+_c \to K^+ \bar K^0 \Lambda$. Thus, we have
\begin{equation}
    \frac{d \Gamma}{dM_{\bar K\Lambda}} \propto P_{\Lambda_c^+ K^+}P_{\bar K\Lambda}(|T_{\bar K\Lambda}|^2+B_1),
\end{equation}
where $P_{\Lambda_c^+ K^+}=\sqrt{\lambda(M^2_{\Lambda_c^+},\,m^2_{K^+},s)}/(2M_{\Lambda_c^+})$ and $P_{\bar K \Lambda}=\sqrt{\lambda(s,m^2_{\bar K}, M^2_\Lambda)}/(2\sqrt{s})$, with $\lambda$ representing the K\"all\'en's Lambda function. The factor that is being summed to the squared amplitude is a background taken as $B_1=1 \times 10^{-3}$.

The data on the invariant mass of $\bar K \Sigma$ are also collected from the decay of $\Lambda^+_c$ measured in Ref.~\cite{Belle:2001hyr}, therefore, we can simply change $\bar K^0 \Lambda$ to $K^-\Sigma^+$ and write
\begin{equation}
     \frac{d \Gamma}{dM_{\bar K\Sigma}} \propto P_{\Lambda_c^+ K^+}P_{\bar K\Sigma}(|T_{\bar K\Sigma}|^2+B_2),
\end{equation}
where the background $B_2=1\times 10^{-4}$ and $P_{\bar K\Sigma}$ is analogous to $P_{\bar K\Lambda}$.

Finally, the invariant mass spectrum of $\pi^+\Xi^-$, provided by the BELLE Collaboration~\cite{Belle:2018lws} has been extracted from the process $\Xi_c^+\to \Xi^-\pi^+\pi^+$. In this case, the presence of  $\Xi(1530)$ can be seen dominantly in the data, which is a spin-3/2 state and is not present in our model. We include it as a Breit-Wigner using $M_{\Xi(1530)}=1532\text{ MeV}$, $\Gamma_{\Xi(1530)}=10 \text{ MeV}$
\begin{equation}
    T_{\Xi(1530)}=\frac{g^2}{\sqrt{s}-M_{\Xi(1530)}+i\Gamma_{\Xi(1530)}/2},
\end{equation}
where $g^2$ is the same as in Ref.~\cite{Khemchandani:2016ftn}, which is approximately $0.47$. 

The invariant mass is calculated as
\begin{equation}
    \frac{d\Gamma}{dM_{\pi\Xi}}\propto P_{\Xi_c^+ \pi^+}P_{\pi\Xi}(|T_{\pi\Xi}|^2+|T_{\Xi(1530)}|^2+B_3),
\end{equation}
where $P_{\Xi_c^+ \pi^+}=\sqrt{\lambda(M^2_{\Xi_c^+},\,m^2_{\pi^+}, s)}/(2M_{\Xi_c^+})$ and the background $B_3=3\times 10^{-3}$.

In order to perform the $\chi^2$-fit, we follow Ref.~\cite{Khemchandani:2018amu}, and calculate  $\chi^2_l$ related to a certain data set as
\begin{equation}\label{chil}
\chi^2_l=\sum_{k=1}^{N^{\text{data}}_l} \frac{(x_{l,k}^{\text{model}}-x^{\text{data}}_{l,k})^2}{\sigma_{l,k}^2},
\end{equation}
where $\sigma_{l,k}$ represents the uncertainty in the data, and $N^{\text{data}}_l$ is the total number of data points in the $l$th set.
We then calculate $\chi^2$ per degree of freedom as a weighted sum of $\chi^2_l/N^\text{data}_l$
\begin{equation}
    \chi^2/{\text{p.d.f}}=W \sum_{l=1}^n \frac{\chi^2_l}{N^{\text{data}}_l},
\end{equation}
where $n$ is the number of distinct data sets considered in the fit, and  
\begin{align}\label{chiweight}
    W=\frac{1}{n}\frac{ \sum\limits_{l=1}^n N^\text{data}_l }{\sum\limits_{l=1}^n N^\text{data}_l-N^\text{para}},
\end{align}
with $N^\text{para}$ being the number of free parameters in the model.
It is useful to mention that $\chi^2/\text{p.d.f}$ provides a quantitative intuition of how much agreement is being reached between the model result and the data. A $\chi^2/\text{p.d.f}$ close to 1 generally indicates a good agreement among the theoretical calculation and the data.

We now present the values of the free parameters, constrained by different data sets, in Table~\ref{para}. 

\begin{table}[h!]
    \centering
    \caption{Values for the free parameters constrained by a fit to a collection of data sets, as shown in Table~\ref{data}. Here $m_V$, and $f_p(f_V)$ represent, respectively, the mass of the vector meson and the decay constant of the pseudoscalar (vector) meson present in a given channel. The values of $f_P$ and $f_V$ are given in Table~\ref{decay}.}  
    \label{para}
    \begin{tabular}{c | c c c}
    \hline \hline
    Parameter & $\qquad \quad$ Fit 1 $\qquad \quad$  &$\qquad \quad$ Fit 2 $\qquad \quad$ & $\qquad \quad$Fit 3 $\qquad \quad$\\ \hline 
    
    $\chi^2/\text{p.d.f}$& $4.51$ & $1.21$ & $8.94$\\
    
    $a_{\pi\Xi}$ & $-1.2230$ &$-2.0362$ & $-1.3911$\\ 
    
    $a_{\eta\Xi}$ & $-2.3155$ & $-3.0103$ & $-3.7624$\\ 
    
    $a_{\bar K\Sigma}$ &  $-2.5021$ & $-1.4647$ & $-2.5056$\\
    
    $a_{\bar K \Lambda}$ & $-3.2314$ & $-2.1205$ & $-3.0199$\\
    
    $a_{\rho\Xi}$ & $-3.1613$ & $-2.1636$ &  $-2.6482$\\ 
    
    $a_{\omega\Xi}$ & $-3.1061$ & $-2.1926$ & $-2.5897$ \\ 
    
    $a_{\phi\Xi}$ & $-2.4546$ & $-3.4180$ &  $-3.0640$\\ 
    
    $a_{\bar K^*\Sigma}$ & $-2.6552$ & $-2.8598$ & $-2.5856$\\ 
    $a_{\bar K^* \Lambda}$ & $-1.9825$ & $-3.4921$ &  $-2.0750$\\ 
    
    $\Lambda$ (MeV) & $954$ & $600$  & $500$\\ 
    
    $q_{max}$ (MeV) \, & $868$ &--& $1000$\\ 
    
    $g_{KR}$ & $0.9647 \times \frac{m_V}{\sqrt {2 \,f_Pf_V}}$ & $0.8478 \times \frac{m_V}{\sqrt {2 \,f_Pf_V}}$  & $0.6298 \times \frac{m_V}{\sqrt {2 \,f_Pf_V}}$
     \\[2mm] \hline \hline   
     \end{tabular}  
\end{table}

Next, we show a comparison of the results acquired in each fit with the available data (summarized in the beginning of this section). It is relevant to mention that we present comparisons not only with the data considered in the fit, but also with the other data sets (which were not considered in the fit).

\subsection{Comparison of model results with the available experimental data}

To begin with, we present the results obtained using the parameters of Fit 1. It is clear from the value $\chi^2/\text{p.d.f}$, presented in the first column of Table~\ref{para}, that the agreement with the data used to perform this first fit is not expected to be satisfactory. In fact, as we already discussed, our attempts of fitting invariant masses and correlation function at the same time were ineffective. 

\begin{figure}[h!]
        \centering        \includegraphics[width=0.9\textwidth]{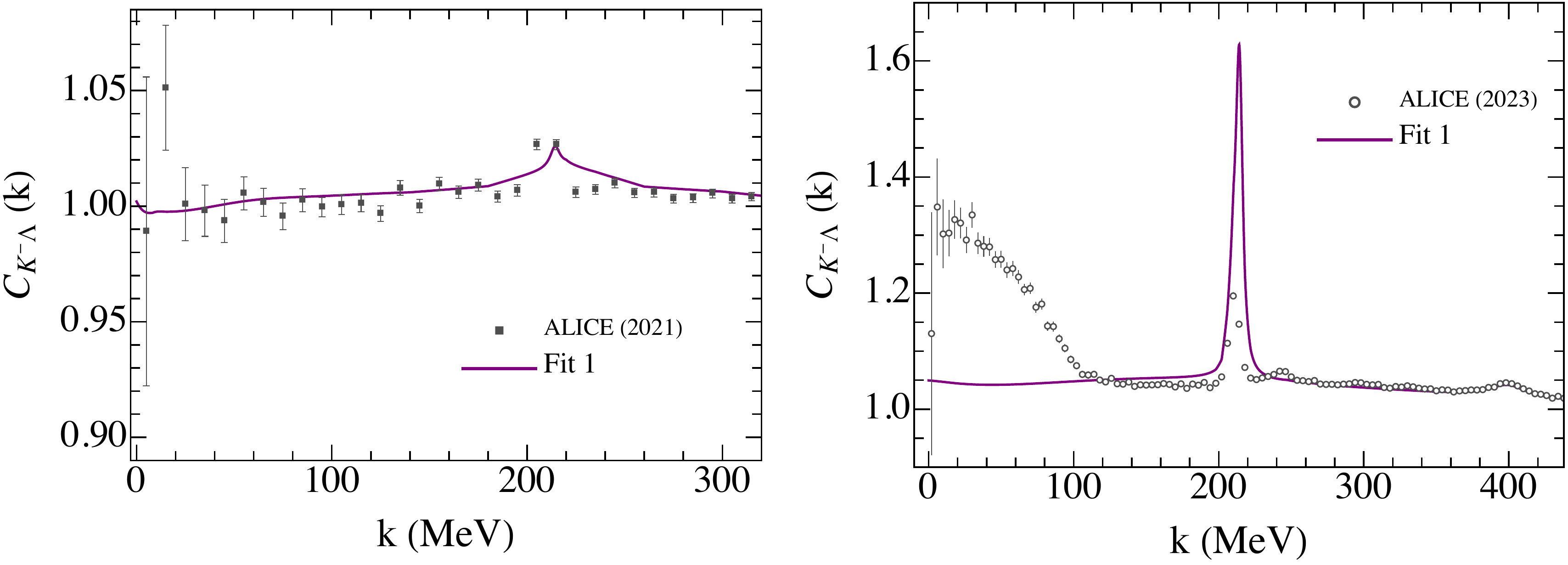}
        \caption{In both panels (left and right), the continuous line represents the result for the $K^-\Lambda$ femtoscopic correlation function obtained within our model using the set of free parameters determined in Fit 1, shown in Table~\ref{para}. In the left (right) panel the filled boxes (open circles) are the experimental data taken from Ref.~\cite{ALICE:2020wvi} (Ref.~\cite{ALICE:2023wjz}) for the correlation function obtained through $Pb\,Pb$ ($p\,p$) collision.}
        \label{CFresult}
\end{figure}

In Fig.~\ref{CFresult},
we show the results on CF obtained within Fit 1. The CF data for $K^-\Lambda$ determined in the $Pb\,Pb$ collision are shown as filled boxes on the left panel, as taken from Ref.~\cite{ALICE:2020wvi}. The thick line represents the results calculated with the parameter set shown in the first column of Table~\ref{para}.  One can clearly see that the description of data, particularly at low momenta, is not satisfactory. It is useful to remind the reader here that precisely this set of data, together with those on the invariant masses, were considered in Fit 1 (see Table~\ref{data}). For completeness, on the right panel, we  show also the results obtained for the  $K^-\Lambda$ CF in the $p\,p$ collision (recall that the source function and the weights are different in $Pb\,Pb$ and $p\,p$ collisions). The experimental data from Ref.~\cite{ALICE:2023wjz} are shown as empty circles on the right panel. The results for this latter case end up with even worse agreement. It is important to clarify that a peak related with the weak decay of $\Omega^-$ is present in the background provided in Ref.~\cite{ALICE:2023wjz}, which is used in our description too [through Eq.~(\ref{exp_def})]. In addition to that we find the generation of a state, in our results obtained in Fit 1, whose mass coincides with that of $\Omega^-$ (see section~\ref{amplitudes} for a detailed discussion on the amplitudes and poles found in each fit).

\begin{figure}[h!]
        \centering
        \includegraphics[width=0.9\textwidth]{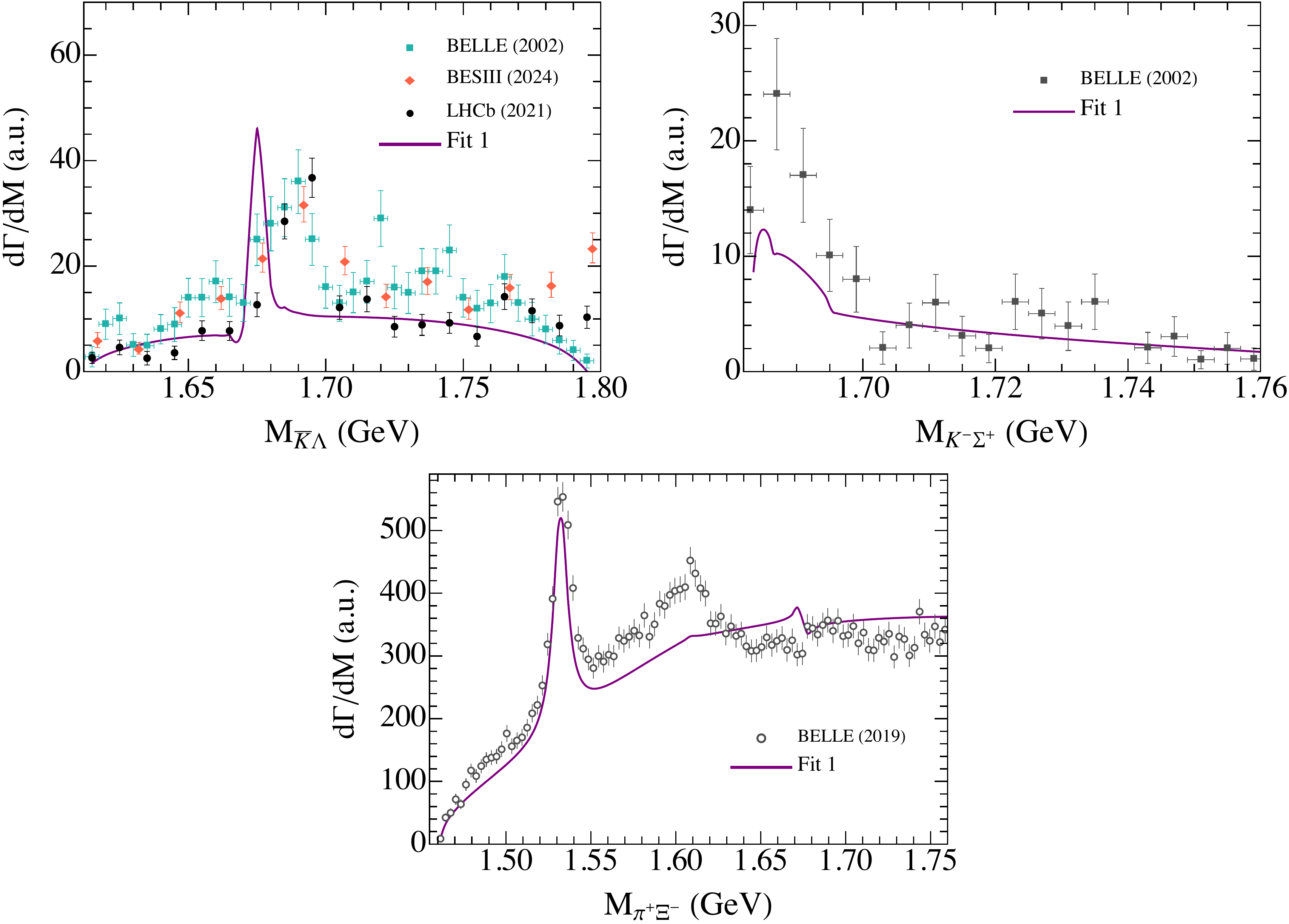}
        \caption{In the upper left (right) panel the  data on the $\bar K \Lambda$ ($K^-\Sigma^+$) invariant mass spectra are taken from  Refs.~\cite{Belle:2001hyr,BESIII:2023mlv,LHCb:2020jpq}. The data on the $\pi^+\Xi^-$ invariant mass shown in the lower panel are taken from Ref.~\cite{Belle:2018lws}. The continuous lines correspond to the results obtained in Fit 1.}
        \label{invmresult}
\end{figure}

Next, in Fig.~\ref{invmresult}, we exhibit a comparison of fit
results with the different invariant mass spectra on which data are available. We would like to recall that the data on the $\bar K \Lambda$, $\bar K \Sigma$ and $\pi\Xi$ invariant masses used as a constraint in Fit 1 are given by the BELLE collaboration, in two different works~\cite{Belle:2001hyr,Belle:2018lws}. In the upper left plot, we show different available data on $\bar K\Lambda$ invariant mass (different data sets have been multiplied by an arbitrary factor for the sake of comparison). It can be seen that the data from BESIII~\cite{BESIII:2023mlv}, BELLE~\cite{Belle:2001hyr} and LHCb~\cite{LHCb:2020jpq} seem to be compatible with each other. 

The results in Fig.~\ref{invmresult} show a poor description of the data on all three systems. These findings led us to consider the $\chi^2$-fitting of the CF and invariant mass spectra separately, as mentioned earlier.

We now discuss the results of these two fits:  Fit 2 and 3. The set of data used to constrain the model parameters in each case is listed in Table~\ref{data} and the resulting values of the parameters are given in Table~\ref{para}.

\begin{figure}[h!]
    \centering
    \includegraphics[width=0.9\textwidth]{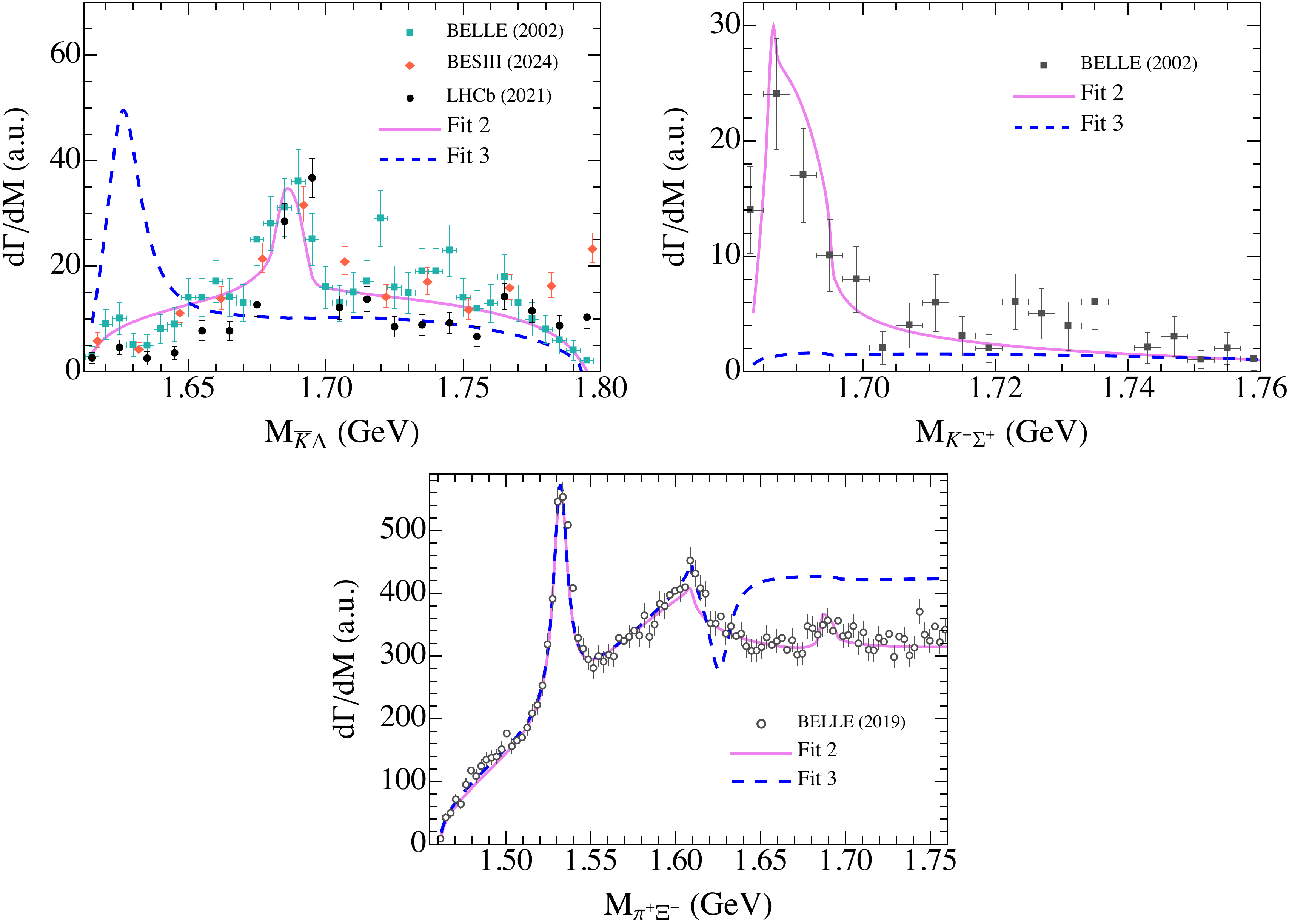}  
     \caption{In the upper left (right) panel the filled boxes are data on the $\bar K^0 \Lambda$ ($K^-\Sigma^+$) invariant mass spectra provided by the BELLE collaboration~\cite{Belle:2001hyr}, the diamond points represent the data reported by BESIII~\cite{BESIII:2023mlv} and the empty circles are the data determined by LHCb~\cite{LHCb:2020jpq}. In the lower panel the open circles points are data on the $\pi^+\Xi^-$ invariant mass, also reported by BELLE collaboration~\cite{Belle:2018lws}. The thick lines are our  result obtained in Fit 2, and the dashed lines show the results of Fit 3.}
    \label{INVM3}
\end{figure}

In Fit 2 the model is constrained to describe the data on the invariant masses of $\bar K\Lambda$, $\bar K \Sigma$ and $\pi\Xi$ from Refs.~\cite{Belle:2001hyr,Belle:2018lws}. The value of $\chi^2/\text{p.d.f}$ in this case is found to be close to 1, so a good description of the data sets considered in the fit is expected. Indeed, one can see that the comparison between the thick lines in Fig.~\ref{INVM3} and the data sets is reasonably good.

\begin{figure}[h!]
    \centering
    \includegraphics[width=0.9\textwidth]{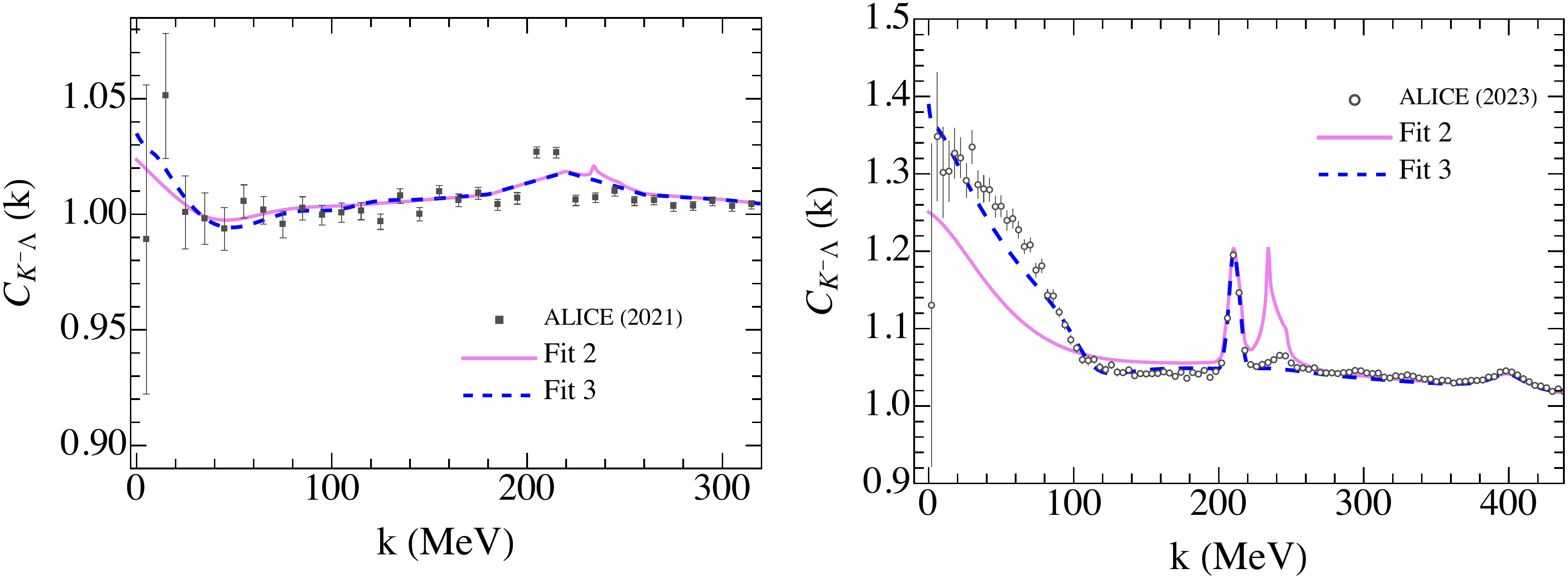}  
     \caption{In both panels (left and right), the thick (dashed) lines represent the result for the $K^-\Lambda$ femtoscopic correlation function obtained within Fit 2 (3). In the left (right) panel the filled boxes (open circles) are the data given by the ALICE Collaboration in Ref.~\cite{ALICE:2020wvi} (Ref.~\cite{ALICE:2023wjz}) for the correlation function obtained through $Pb\,Pb$ ($p\,p$) collision.}
    \label{CF3}
\end{figure}

One would expect that the same parameters should be able to describe the data on the correlation function too. We calculate, with the parameters of Fit 2 and considering $q_{max}=800 \text{ MeV}$, the $K^-\Lambda$ CF and compare with the two different measurements of the ALICE collaboration~\cite{ALICE:2020wvi,ALICE:2023wjz}. The results are presented in Fig.~\ref{CF3}. It is possible to notice some improvement in the description of data in the low momentum region. However, the results are still not in agreement with the data.

Finally, Fit 3 was performed by constraining the free parameters of our model to  describe  the $K^-\Lambda$ CF extracted in the collision $p\,p$. The results are presented in Fig.~\ref{CF3} as dashed lines. A value of $\chi^2/\text{p.d.f}$ far from unity is obtained in this case (as shown in the third column of Table~\ref{para}), although the fit may not appear very distant from the data points. In fact, we find $\sum\limits_i\left(x^{\text{model}}_i-x^{\text{data}}_i\right)^2<0.1$, but  $\chi^2$ still becomes large because the error bars [$\sigma$ in Eq.~(\ref{chil})] on the data are very small. It is relevant to recall that $\Xi(1690)$ is included in the background in Ref.~\cite{ALICE:2023wjz} using a Breit-Wigner form. However, we removed it from the background as mentioned earlier. We do this because we find this state to get generated from the dynamics when parameter sets 1 and 2 are used. Hence, the missing signal in the results shown with the dashed line is in agreement with the analysis of ALICE.

Now one would expect that the same parameters that describe the CF, at least in the low momentum region, should be able to describe the data on the $\bar K\Lambda$ invariant mass spectra. We show the results of the invariant masses of $\pi\Xi$, $\bar K \Lambda$ and $\bar K \Sigma$,  obtained using the parameters of Fit 3, as the dashed lines in Fig.~\ref{INVM3}. It is evident that the dashed lines do not describe the $\bar K \Lambda$ and $\bar K \Sigma$ data (also $\pi \Sigma$, at energies where $\bar K \Lambda$ and $\bar K \Sigma$ thresholds are open). Our findings indicate that either something is missing in the model, particularly in relation with the calculation of the correlation function, or that the data sets on CF and invariant mass spectra may have some incompatibility.

Further, we evaluate the scattering length for the $K^-\Lambda$ system, motivated by the results reported by the ALICE collaboration~\cite{ALICE:2020wvi, ALICE:2023wjz}. In both references, the authors extracted the values of the scattering length through the femtoscopic CF. The scattering length, $a_{MB}$, of a meson($M$)-baryon($B$) system, in our model, is calculated as 
\begin{equation}
    a_{MB}=-\frac{T_{MB}(\sqrt {s_{thr}})}{\sqrt {s_{thr}}}\frac{M_B}{4\pi},
\end{equation}
where $\sqrt{s_{thr}}=M_B+m_M$ is the threshold of the channel.

We give the results of $a_{MB}$ for the $K^-\Lambda$ system in Table~\ref{scat}, obtained by using each of the three sets of parameters listed in Table~\ref{para}, together with the values determined experimentally~\cite{ALICE:2020wvi,ALICE:2023wjz}.
\begin{table}[h!]
    \centering
    \caption{Values of the scattering length for the system $K^-\Lambda$ obtained using the parameters constrained by Fit 1, Fit 2 and Fit 3, using the data as in Table~\ref{data}.}
    \begin{tabular}{c | c c c c c}
    \hline \hline
     &$\quad$ Fit 1 $\quad$ &$\quad$ Fit 2 $\quad$ & $\quad$Fit 3 $\quad$ & $\quad$ ALICE(2021)~\cite{ALICE:2020wvi} $\quad$ & $\quad$ ALICE(2023)~\cite{ALICE:2023wjz} $\quad$\\ \hline
     Re$\{ a_0\} [\text{fm}]$ &$-0.07$& $0.26$ &  $0.37$ & $0.27 \pm 0.127$ & $0.33 \pm 0.05$ \\

     Im$\{ a_0\}  [\text{fm}]$ &$0.07$& $0.42$ & $0.53$ & $0.40\pm 0.117$ & $0.46 \pm 0.05$
     \\ \hline \hline
     
    \end{tabular}
    \label{scat}
\end{table}
The data provided by the ALICE Collaboration are  compatible with each other, when considering error bars. The results deduced in Fit 2 and Fit 3 show a good agreement with the data, but the Fit 1 results are not compatible with the data.

To complete the discussions of this section, we now show the modulus squared amplitudes for the different channels, as obtained in the three different fits. We also tabulate the properties of the poles found in each case.

\subsection{ Scattering amplitudes and resonance poles}\label{amplitudes}

We find it useful to present the modulus squared scattering amplitudes for  $\pi\Xi$, $\bar K \Sigma$ and $\bar K \Lambda$ systems in spin and isospin 1/2, since those are precisely the systems for which a comparison with the experimental data has been made in the previous  section. 

\begin{figure}[h!]
        \centering
        \includegraphics[width=0.85\textwidth]{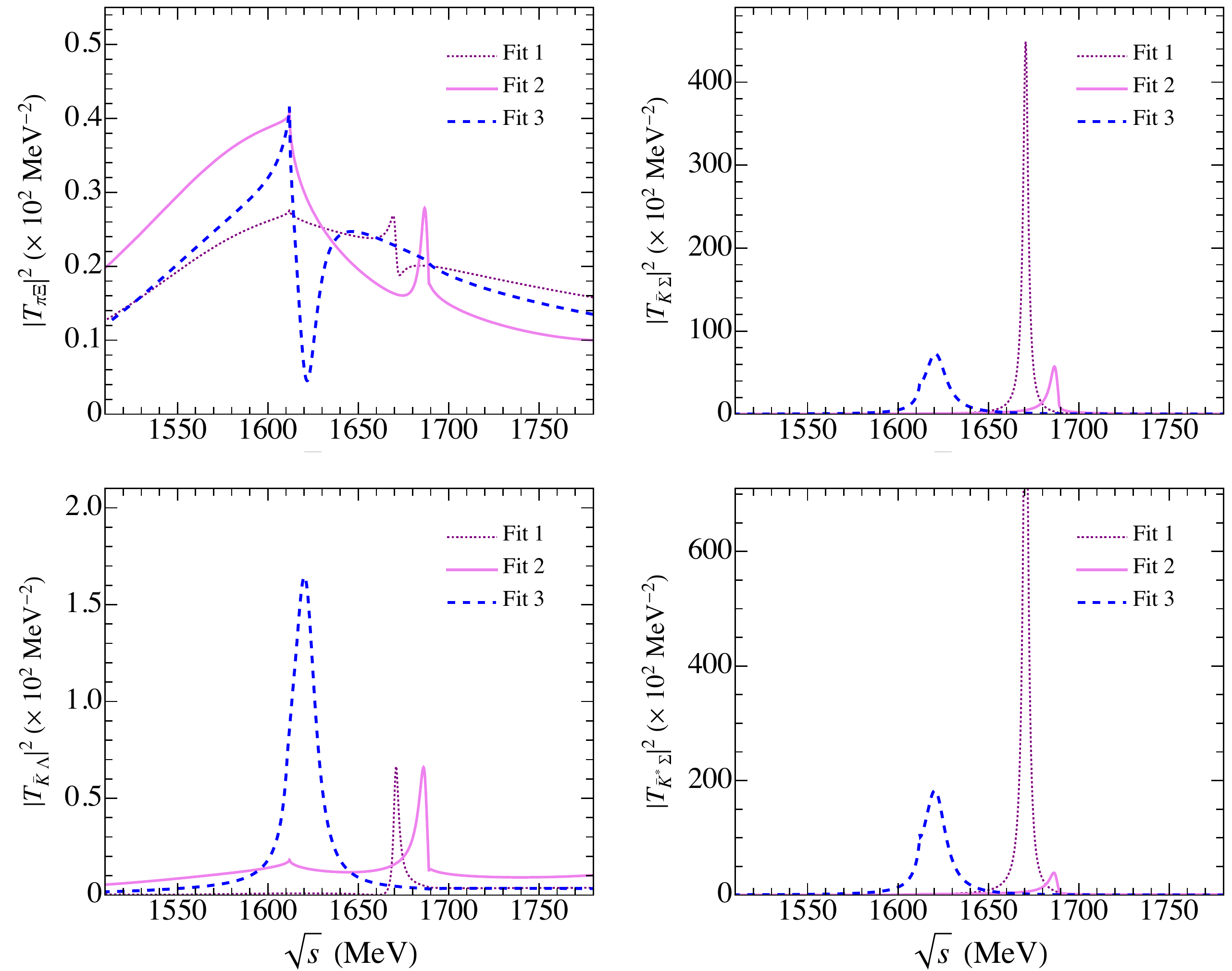}
        \caption{Squared amplitudes of the three ($\pi\Xi$, $\bar K \Lambda$ and $\bar K \Sigma$) pseudoscalar-baryon systems for which  experimental data are available. For completeness. We also present a vector-baryon amplitude (for $\bar K^*\Sigma$).  The  dotted, thick and the dashed line are the results obtained in Fits 1, 2 and 3, respectively.}
        \label{AMPnew}
\end{figure}

In Fig.~\ref{AMPnew}, we show the modulus squared amplitudes for the pseudoscalar-baryon channels obtained with the parameter sets for the three fits discussed in this manuscript. We show the results for one of the vector-baryon systems as well, for the sake of completeness.
It is worth focusing on the results obtained in Fits 2 and 3, since the results in this case have some agreement with at least a partial set of the experimental data. 

We can start by discussing that a prominent structure  near 1600 MeV is seen in the results shown with a dashed line, which corresponds to Fit~3. We can identify this state with $\Xi(1620)$. In the $\pi \Xi$ amplitude, the signal seems to get distorted by the opening of the $\bar K\Lambda$ threshold.  In this case, no clear evidence is found for $\Xi(1690)$. It is important to recall here that Fit 3 describes better the correlation function data, but  leads to results in disagreement with data on the invariant mass spectra of different systems (see Figs.~\ref{INVM3} and \ref{CF3}). 

The  signal seen near 1600 MeV, in the Fit 3 results, is lost in the amplitudes obtained in Fit 2, while a peak structure near 1690 MeV appears. We recall that Fit 2 describes well the data on the invariant mass spectra but not the data on the correlation function (see Figs.~\ref{INVM3} and \ref{CF3}).

To draw further conclusions, it is useful to provide information on the poles found in each of the three fits discussed above. To search for poles  we must continue the amplitudes analytically to the complex plane. To do so, we write the loop function $G_j$ in Eq.~(\ref{loop})  above the threshold of the $j$th channel as
\begin{equation}
    G^{\text{II}}_j(\sqrt s)=G^I_j(\sqrt s)-2i\text{Im}\{G^I_j(\sqrt s)\},\label{GII}
\end{equation}
where the superscript II (I) refers to the second (first) Riemann sheet. The imaginary part of the loop function is constrained by the unitarity of the $S-$matrix
\begin{align}
   \text{Im}\{G^I_j(\sqrt s)\}=- \frac{M_j~q_j}{4\pi \sqrt{s}},
\end{align}
where $M_j$ and $q_j$ stand for the mass of the baryon in $j$th channel and the center of mass momentum of the system.

For each of the sets of parameters constrained by the fits discussed in the last section we find three different poles, as summarized in Table~\ref{pole}. 
\begin{table}[h!]
\centering
\scriptsize
\caption{Positions for the poles obtained in the amplitudes shown in Fig.~\ref{AMPnew}, when continued analytically to the complex energy plane.}
\begin{tabular}{c|ccc|ccc|ccc}
\hline\hline
& \multicolumn{3}{c|}{Fit 1} & \multicolumn{3}{c|}{Fit 2} & \multicolumn{3}{c}{Fit 3} \\
& $\,$ Pole 1 $\,$ & $\,$ Pole 2 $\,$  & $\,$  Pole 3 $\,$ 
& $\,$ Pole 1 $\,$ & $\,$ Pole 2 $\,$  & $\,$  Pole 3 $\,$ 
& $\,$ Pole 1 $\,$ & $\,$ Pole 2 $\,$  & $\,$  Pole 3 $\,$ \\
\hline
Mass (MeV)
& $1557.5$ & $1670.0$ & $1960.0$
& $1576.0$ & $1687.0$ & $2086.5$
& $1561.0$ & $1620.0$ & $2047.0$ \\
Width (MeV)
& $251.0$ & $3.0$ & $35.0$
& $241.0$ & $6.0$ & $16.0$
& $265.0$ & $15.0$ & $50.0$ \\
Correspondence with a known state& -- & $\Xi(1690)$ & $\Xi(1950)$
& -- & $\Xi(1690)$ & --
& -- & $\Xi(1620)$ & -- \\\hline\hline
\end{tabular}\label{pole}
\end{table}

It can be seen that either a state related to $\Xi(1690)$ or  to $\Xi(1620)$ is found but the two do not appear simultaneously. It is not surprising that the pole related to $\Xi(1620)$ has properties in agreement with those determined by the ALICE collaboration~\cite{ALICE:2023wjz}, since Fit 3 best explains the data on the correlation function. Once again, the same amplitudes, however, fail to describe the data on the invariant mass spectra. The fit that better describes the invariant mass spectra~\cite{Belle:2001hyr,Belle:2018lws,LHCb:2020jpq,BESIII:2015dvj,BESIII:2023mlv}, on the other hand, consists of the pole related to $\Xi(1690)$ and not the state with properties determined by ALICE. 

Besides the cascade states in the 1620-1690 MeV region, interestingly, a wide pole always appears in the range 1557-1576~MeV. A state with similar properties seems to appear whenever Weinberg-Tomozawa interactions based on lowest order $\chi$PT are considered in meson-baryon systems~\cite{Ramos:2002xh,Gamermann:2011mq}.
Such a pole cannot be associated to any known states listed in Ref.~\cite{PDG2024} as indicated in Table~\ref{pole}. It is possible that a state having a width of 250 MeV, if exists, may not be easy to observe since this energy region is dominated by the presence of $\Xi(1530)$ in the experimental data.

\begin{figure}[h!]
        \centering
\includegraphics[width=0.42\textwidth]{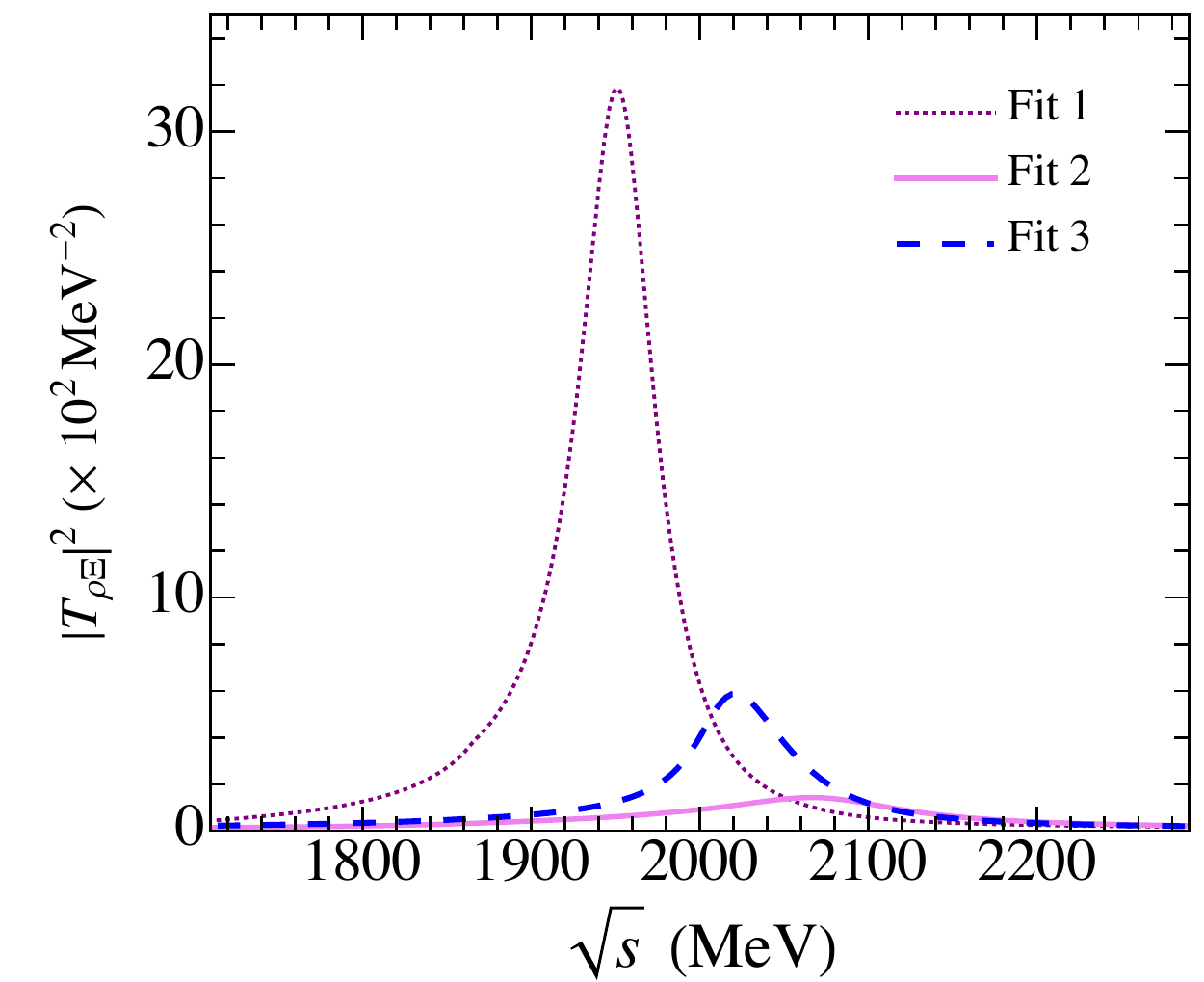}
        \caption{Modulus squared amplitude of the $\rho \Xi$ channel in isospin 1/2, spin 1/2 configuration, showing the presence of the third pole listed in Table~\ref{pole}.}
        \label{amprhoxi}
\end{figure}

Also, a pole at higher energies (between 1960-2090 MeV) appears in all cases. It is important to mention that this pole lies in the region beyond the energies related to the data considered  to fit the parameters of the model. We can identify the pole found in Fit~1 with $\Xi(1950)$. To see the effect of the presence of this state on the real axis, we show in Fig.~\ref{amprhoxi} the $\rho\Xi$ modulus squared amplitude obtained for isospin $1/2$ and spin $1/2$. This energy region is not shown in Fig.~\ref{AMPnew} for clarity. A clear signal is seen in the energy region 1950-2050 MeV. 
Spin degenerate states in this energy region have also been found in the models of Refs.~\cite{Oset:2010tof,Gamermann:2011mq} based on Weinberg-Tomozawa interactions. Specifically, the authors of Ref.~\cite{Oset:2010tof} report two poles at $2039- i 67$ MeV and $2082- i 31$ MeV, with the former one coupling strongly to $\rho \Xi$ and the latter one coupling more to $\bar K^*\Sigma$. Similar results have been found in Ref.~\cite{Gamermann:2011mq}, where poles at $2037- i 24$ MeV and $2094- i 59$ MeV have been listed. From the experimental side, the situation in this energy region is not very clear. The particle data group~\cite{PDG2024} lists two cascades: $\Xi(1950)$ and $\Xi(2030)$ but spin-parity of both states are still not established. 
 In fact, data at these energies are available on the invariant mass spectrum of $\bar K\Lambda$ from LHCb but it is hard to draw any conclusions since the data have very large uncertainties in this energy region\cite{LHCb:2020jpq}. We hope that data with better statistics are made available in the future to clarify the status of doubly strange states with mass values around 2 GeV.

It might be also useful  to present the couplings of the three poles to each one of the nine channels, which may help in understanding the nature of the associated states. The couplings are evaluated through the calculation of the residues of the scattering amplitude at the pole position. The values of the couplings $g_j$ are summarized in Tables~\ref{res} and \ref{res2}.
\begin{table}[h!]
\centering
\scriptsize
\caption{Residues for each pole with the set of parameters of Fit 1.}
\begin{tabular}{l|ccc}
\hline\hline
& \multicolumn{3}{c}{Fit 1} \\
& $\qquad$ Pole 1 $\qquad$ & $\qquad$ Pole 2 $\qquad$  &  $\qquad$  Pole 3 $\qquad$   \\
\hline
$g_{\pi\Xi}$
& $-1.55+i1.61$ & $-0.10-i0.05$ & $0.13-i0.17$ \\
$g_{\eta\Xi}$
& $-0.61+i1.02$ & $1.42+i0.02$ & $-0.46+i0.06$ \\
$g_{\bar K \Sigma}$
& $0.33-i1.00$ & $1.91+0.00$ & $0.46-i0.24$ \\
$g_{\bar K \Lambda}$
& $1.38-i0.49$ & $-0.35+i0.08$ & $0.36-i0.18$ \\
$g_{\rho\Xi}$
& $-0.96-i0.66$ & $0.72-i0.09$ & $3.53+i0.15$ \\
$g_{\omega\Xi}$
& $0.18+i0.23$ & $-1.00+i0.00$ & $-0.36-i0.20$ \\
$g_{\phi\Xi}$
& $-0.28-i.034$ & $1.63-i0.00$ & $0.44+i0.28$ \\
$g_{\bar K^*\Sigma}$
& $-0.40-i0.21$ & $-2.34-i0.06$ & $1.20+i0.32$ \\
$g_{\bar K^*\Lambda}$
& $0.30-i0.68$ & $0.32-i0.04$ & $-0.46+i0.36$  \\
\hline\hline
\end{tabular}\label{res}
\end{table}
\begin{table}[h!]
\centering
\scriptsize
\caption{Residues for each pole with the set of parameters of Fit 2 and Fit 3.}
\begin{tabular}{l|ccc|ccc}
\hline\hline
& \multicolumn{3}{c|}{Fit 2} & \multicolumn{3}{c}{Fit 3} \\
& $\qquad$ Pole 1 $\qquad$ & $\qquad$ Pole 2 $\qquad$  &  $\qquad$  Pole 3 $\qquad$  
& $\qquad$ Pole 1 $\qquad$ & $\qquad$ Pole 2 $\qquad$  &  $\qquad$  Pole 3 $\qquad$  \\
\hline
$g_{\pi\Xi}$
& $-2.37+i1.25$ & $-0.23-i3.00$ & $0.09-i0.09$
& $-1.77+i2.00$ & $0.15+i0.53$ & $-0.05-i0.24$ \\
$g_{\eta\Xi}$
& $0.13+i1.05$ & $0.86+i0.42$ & $-0.22-i0.02$
& $-0.29+i0.78$ & $1.69+i0.12$ & $-0.29+i0.15$ \\
$g_{\bar K \Sigma}$
& $0.78-i0.5$ & $1.53+i0.33$ & $0.18-i0.33$
& $0.80-i0.31$ & $2.53-i.022$ & $0.27-i0.37$ \\
$g_{\bar K \Lambda}$
& $1.59-i1.11$ & $-0.40-i0.20$ & $0.14-i0.19$
& $1.41-i0.96$ & $-0.90-i0.18$ & $0.27-i0.37$ \\
$g_{\rho\Xi}$
& $-0.96-i0.25$ & $-0.01-i0.05$ & $1.53+i0.47$
& $-0.62-i0.25$ & $0.15-0.04$ & $3.16+i0.21$ \\
$g_{\omega\Xi}$
& $0.43+i0.07$ & $-0.32+i0.04$ & $-0.09+i0.07$
& $-0.02+i0.04$ & $-0.75+i0.05$ & $-0.03+i0.31$ \\
$g_{\phi\Xi}$
& $-0.68-i0.10$ & $0.55-i0.05$ & $0.19-i0.09$
& $0.04-0.07$ & $1.24-i0.08$ & $0.11-i0.46$ \\
$g_{\bar K^*\Sigma}$
& $-1.52-i1.11$ & $-1.23-i0.70$ & $0.21-i0.03$
& $-0.48-i0.84$ & $-3.20-i0.03$ & $0.36-i0.43$ \\
$g_{\bar K^*\Lambda}$
& $-1.22-i0.64$ & $0.24-i0.33$ & $-0.25-i0.20$
& $0.01-i0.44$ & $0.26-i0.17$ & $-1.03-i0.54$ \\
\hline\hline
\end{tabular}\label{res2}
\end{table}
We can see from the couplings obtained within the three sets of parameters that the first pole always couples more to $\pi\Xi$ and $\bar K \Lambda$, whereas the second pole always couples strongly to $\eta \Xi$ and $\bar K \Sigma$. As for pole 3, it can be seen that it is related mostly to the $\rho\Xi$ channel.

\section{Summary}
The findings of our work can be summarized as follows:
\begin{itemize}
    \item We find that it is not possible to fit the data on the correlation function of $\bar K\Lambda$ and the invariant mass spectra of the same and other channels simultaneously.
    \item An attempt to fit the  correlation function leads to an amplitude that can describe roughly the behavior of the experimental data. However, the same amplitude gives rise to an invariant mass spectrum of the same system, which is in complete disagreement with the related experimental data.
    \item A separate attempt made to fit the data on the  invariant mass spectra of various pseudoscalar-baryon systems with strangeness $-2$ gives a good $\chi^2$ value (all data are reproduced satisfactorily). But the correlation function calculated with the same $\bar K\Lambda$ amplitude completely disagrees with the CF data.
    \item In both cases: where a fit is made only to the correlation function or to the invariant mass spectra, the scattering lengths obtained are in good agreement with the ALICE data~\cite{ALICE:2020wvi,ALICE:2023wjz}.
\end{itemize}
We hope that studies with other models or other considerations, which determine observables related to  the available data on strangeness $-2$ can shed more light on this issue.
\section{Acknowledgements}
The authors are thankful to Philipp Gubler for giving very useful comments and to Rafael Coutinho, Mizuki Sumihama, and Hongxun Yang for providing experimental data used in this work.
This study was financed in part by the Coordena\c c\~ao de Aperfei\c coamento de Pessoal de N\'ivel Superior – Brasil (CAPES) – Finance Code 001.
The partial support from other Brazilian agencies is also gratefully acknowledged. We thank CNPq (K.P.K: Grants No. 407437/ 2023-1 and No. 306461/2023-4; A.M.T: Grant No. 304510/2023-8; L.M.A.: Grants No. 400215/2022-5, 308299/2023-0, 402942/2024-8), and CNPq/FAPERJ under the Project INCT-F\'{\i}sica Nuclear e Aplica\c c\~oes (Contract No. 408419/2024-5). A.M.T thanks FAPESP, project 2025/20068-6. L.S.G. is partly supported by the National Natural Science Foundation of China under Grants No.W2543006 and No.12435007.

\section{Appendix} 
In this Appendix we provide the tables containing the SU(3) coefficients needed to evaluate the lowest order scattering amplitudes given in Eqs.~(\ref{sijeq}), (\ref{uijeq}), (\ref{VB:t}), (\ref{VB:CT}), (\ref{VB:su}) and (\ref{krolleq}).
\begin{table}[h!]
    \centering
    \scriptsize
    \caption{$S_{ij}$ coefficients used in Eq.~(\ref{sijeq}).}
    \begin{tabular}{c | l | l | l | l }
    \hline  \hline
    & $\pi\Xi$ & $\eta\Xi$ & $\bar K \Sigma$ & $\bar K \Lambda$ \\ \hline 
        
    $\pi\Xi$ & \begin{tabular}{c | c}
    $\Xi$ & $3\,(D^\prime-F^\prime)^2$\\
    $N$ & 0\\
    $\Sigma$ & 0\\
    $\Lambda$ & 0 \\ \\
    \end{tabular} &

    \begin{tabular}{c | c}
    $\Xi$ & $-(D^\prime-F^\prime)(D^\prime+3\,F^\prime)$\\
    $N$ & 0\\
    $\Sigma$ & 0\\
    $\Lambda$ & 0 \\ \\
    \end{tabular} &

    \begin{tabular}{c | c}
    $\Xi$ & $-3\,(D^\prime-F^\prime)(D^\prime+F^\prime)$\\
    $N$ & 0\\
    $\Sigma$ & 0\\
    $\Lambda$ & 0\\ \\
    \end{tabular} &

    \begin{tabular}{c | c}
    $\Xi$ & $-(D^\prime-F^\prime)(D^\prime-3\,F^\prime)$\\
    $N$ & 0\\
    $\Sigma$ & 0\\
    $\Lambda$ & 0\\ \\
    \end{tabular} \\

    $\eta\Xi$ & \begin{tabular}{c | c}
    $\Xi$ & $-(D^\prime-F^\prime)(D^\prime+3\,F^\prime)$ \\
    $N$ & 0 \\
    $\Sigma$ & 0 \\
    $\Lambda$ & 0 \\ \\
    \end{tabular} &

    \begin{tabular}{c | c}
    $\Xi$ & $\frac{1}{3}\,(D^\prime+3F^\prime)^2$\\
    $N$ & 0\\
    $\Sigma$ & 0 \\
    $\Lambda$ & 0 \\ \\
    \end{tabular} &

    \begin{tabular}{c | c}
    $\Xi$ & $(D^\prime+F^\prime)(D^\prime+3\,F^\prime)$\\
    $N$ & 0\\
    $\Sigma$ & 0\\
    $\Lambda$ & 0\\ \\
    \end{tabular} &

    \begin{tabular}{c | c}
    $\Xi$ & $\frac{1}{3}(D^\prime-3F^\prime)(D^\prime+3\,F^\prime)$\\
    $N$ & 0\\
    $\Sigma$ & 0\\
    $\Lambda$ & 0\\ \\
    \end{tabular} \\

    $\bar K\Sigma$ & \begin{tabular}{c | c}
    $\Xi$ & $-3\,(D^\prime-F^\prime)(D^\prime+F^\prime)$ \\
    $N$ & 0 \\
    $\Sigma$ & 0 \\
    $\Lambda$ & 0 \\ \\
    \end{tabular} &

    \begin{tabular}{c | c}
    $\Xi$ & $(D^\prime+F^\prime)(D^\prime+3\,F^\prime)$\\
    $N$ & 0\\
    $\Sigma$ & 0 \\
    $\Lambda$ & 0 \\ \\
    \end{tabular} &

    \begin{tabular}{c | c}
    $\Xi$ & $3\,(D^\prime+F^\prime)^2$\\
    $N$ & 0\\
    $\Sigma$ & 0\\
    $\Lambda$ & 0\\ \\
    \end{tabular} &

    \begin{tabular}{c | c}
    $\Xi$ & $(D^\prime-3\,F^\prime)(D^\prime+F^\prime)$\\
    $N$ & 0\\
    $\Sigma$ & 0\\
    $\Lambda$ & 0\\ \\
    \end{tabular} \\

    $\bar K\Lambda$ & \begin{tabular}{c | c}
    $\Xi$ & $-(D^\prime-F^\prime)(D^\prime-3\,F^\prime)$ \\
    $N$ & 0 \\
    $\Sigma$ & 0 \\
    $\Lambda$ & 0 \\ \\
    \end{tabular} &

    \begin{tabular}{c | c}
    $\Xi$ & $\frac{1}{3}(D^\prime-3F^\prime)(D^\prime+3\,F^\prime)$\\
    $N$ & 0\\
    $\Sigma$ & 0 \\
    $\Lambda$ & 0 \\ \\
    \end{tabular} &

    \begin{tabular}{c | c}
    $\Xi$ & $(D^\prime-3\,F^\prime)(D^\prime+F^\prime)$\\
    $N$ & 0\\
    $\Sigma$ & 0\\
    $\Lambda$ & 0\\ \\
    \end{tabular} &

    \begin{tabular}{c | c}
    $\Xi$ & $\frac{1}{3}(D^\prime-3F^\prime)^2$\\
    $N$ & 0\\
    $\Sigma$ & 0\\
    $\Lambda$ & 0\\ \\
    \end{tabular} \\
    \hline \hline    
    \end{tabular}
    \label{sijk}
\end{table}
\begin{table}[h!]
    \centering
    \scriptsize
    \caption{$U_{ij,\text{PB}}^{k}$ coefficients used in Eq.~(\ref{uijeq}).}
    \begin{tabular}{c | l | l | l | l }
    \hline  \hline
    & $\pi\Xi$ & $\eta\Xi$ & $\bar K \Sigma$ & $\bar K \Lambda$ \\ \hline 
        
    $\pi\Xi$ & \begin{tabular}{c | c}
    $\Xi$ & $-(D^\prime-F^\prime)^2$\\
    $N$ & 0\\
    $\Sigma$ & 0\\
    $\Lambda$ & 0 \\ \\
    \end{tabular} &

    \begin{tabular}{c | c}
    $\Xi$ & $-(D^\prime-F^\prime)(D^\prime+3\,F^\prime)$\\
    $N$ & 0\\
    $\Sigma$ & 0\\
    $\Lambda$ & 0 \\ \\
    \end{tabular} &

    \begin{tabular}{c | c}
    $\Xi$ & 0\\
    $N$ & 0\\
    $\Sigma$ & $4\,F^\prime(D^\prime+F^\prime)$\\
    $\Lambda$ & $\frac{2}{3}\,D^\prime(D^\prime-3\,F^\prime)$\\ \\
    \end{tabular} &

    \begin{tabular}{c | c}
    $\Xi$ & 0\\
    $N$ & 0\\
    $\Sigma$ & $2\,D^\prime(D^\prime+F^\prime)$\\
    $\Lambda$ & 0\\ \\
    \end{tabular} \\

    $\eta\Xi$ & \begin{tabular}{c | c}
    $\Xi$ & $-(D^\prime-F^\prime)(D^\prime+3\,F^\prime)$ \\
    $N$ & 0 \\
    $\Sigma$ & 0 \\
    $\Lambda$ & 0 \\ \\
    \end{tabular} &

    \begin{tabular}{c | c}
    $\Xi$ & $\frac{1}{3}\,(D^\prime+3F^\prime)^2$\\
    $N$ & 0\\
    $\Sigma$ & 0 \\
    $\Lambda$ & 0 \\ \\
    \end{tabular} &

    \begin{tabular}{c | c}
    $\Xi$ & 0\\
    $N$ & 0\\
    $\Sigma$ & $-2\,D^\prime(D^\prime+F^\prime)$\\
    $\Lambda$ & 0\\ \\
    \end{tabular} &

    \begin{tabular}{c | c}
    $\Xi$ & 0\\
    $N$ & 0\\
    $\Sigma$ & 0\\
    $\Lambda$ & $\frac{2}{3}\,D^\prime(D^\prime-3F^\prime)$\\ \\
    \end{tabular} \\

    $\bar K\Sigma$ & \begin{tabular}{c | c}
    $\Xi$ & 0 \\
    $N$ & 0 \\
    $\Sigma$ & $4\,F^\prime(D^\prime+F^\prime)$\\
    $\Lambda$ & $\frac{2}{3}\,D^\prime(D^\prime-3\,F^\prime)$ \\ \\
    \end{tabular} &

    \begin{tabular}{c | c}
    $\Xi$ & 0\\
    $N$ & 0\\
    $\Sigma$ & $-2\,D^\prime(D^\prime+F^\prime)$ \\
    $\Lambda$ & 0 \\ \\
    \end{tabular} &

    \begin{tabular}{c | c}
    $\Xi$ & 0\\
    $N$ & $-(D^\prime-F^\prime)^2$\\
    $\Sigma$ & 0\\
    $\Lambda$ & 0\\ \\
    \end{tabular} &

    \begin{tabular}{c | c}
    $\Xi$ & 0\\
    $N$ & $(D^\prime-F^\prime)(D^\prime+3\,F^\prime)$\\
    $\Sigma$ & 0\\
    $\Lambda$ & 0\\ \\
    \end{tabular} \\

    $\bar K\Lambda$ & \begin{tabular}{c | c}
    $\Xi$ & 0 \\
    $N$ & 0 \\
    $\Sigma$ & $2\,D^\prime(D^\prime+F^\prime)$ \\
    $\Lambda$ & 0 \\ \\
    \end{tabular} &

    \begin{tabular}{c | c}
    $\Xi$ & 0\\
    $N$ & 0\\
    $\Sigma$ & 0 \\
    $\Lambda$ & $\frac{2}{3}\,D^\prime(D^\prime-3F^\prime)$ \\ \\
    \end{tabular} &

    \begin{tabular}{c | c}
    $\Xi$ & 0\\
    $N$ & $(D^\prime-F^\prime)(D^\prime+3\,F^\prime)$\\
    $\Sigma$ & 0\\
    $\Lambda$ & 0\\ \\
    \end{tabular} &

    \begin{tabular}{c | c}
    $\Xi$ & 0\\
    $N$ & $\frac{1}{3}\,(D^\prime+3\,F^\prime)$\\
    $\Sigma$ & 0\\
    $\Lambda$ & 0\\ \\
    \end{tabular} \\
    \hline \hline    
    \end{tabular}
    \label{uijk}
\end{table}

\begin{table}[h!]
    \centering
    \caption{$\mathcal{A}_{ij}$ coefficients used in Eq.~(\ref{VB:t}).}
    \begin{tabular}{c|p{2.6cm} p{2.6cm} p{2.6cm} p{2.6cm} p{2.6cm}}
    \hline \hline
     & $\rho\Xi$ & $\omega\Xi$ & $\phi\Xi$ & $\bar K^*\Sigma$ & $\bar K^*\Lambda$ \\ \hline

     $\rho\Xi$ & $2$ & $0$ & $0$ & $-\frac{1}{2}$ & $-\frac{3}{2}$\\

     $\omega\Xi$ & $0$ & $0$ & $0$ & $\frac{\sqrt{3}}{2}$ & $-\frac{\sqrt{3}}{2}$ \\ 

     $\phi\Xi$ & $0$ & $0$ & $0$ & $-\sqrt\frac{3}{2}$ & $\sqrt\frac{3}{2}$\\

     $\bar K^*\Sigma$ & $-\frac{1}{2}$ & $\frac{\sqrt{3}}{2}$ & $-\sqrt\frac{3}{2}$ & $2$ & $0$ \\

     $\bar K^*\Lambda$ & $-\frac{3}{2}$ & $-\frac{\sqrt{3}}{2}$ & $\sqrt\frac{3}{2}$ & $0$ & $0$ \\

     \hline \hline
     
    \end{tabular}    
    \label{aij}
\end{table}

\begin{table}[h!]
    \centering
    \caption{$\mathcal{B}_{ij}$ coefficients used in Eq.~(\ref{VB:CT})}
    \begin{tabular}{c|p{2.6cm} p{2.6cm} p{2.6cm} p{2.6cm} p{2.6cm}}
    \hline \hline
     & $\rho\Xi$ & $\omega\Xi$ & $\phi\Xi$ & $\bar K^*\Sigma$ & $\bar K^*\Lambda$ \\ \hline

     $\rho\Xi$ & $F-D$ & $0$ & $0$ & $-\frac{(D+F)}{4}$ & $\frac{(D-3\,F)}{4}$\\

     $\omega\Xi$ & $0$ & $0$ & $0$ & $\frac{\sqrt{3}(D+F)}{4}$ & $\frac{1}{4}\frac{(D-3\,F)}{\sqrt{3}}$ \\ 

     $\phi\Xi$ & $0$ & $0$ & $0$ & $-\frac{\sqrt{3}(D+F)}{2\sqrt{2}}$ & $-\frac{(D-3\,F)}{2\,\sqrt{6}}$\\

     $\bar K^*\Sigma$ & $-\frac{(D+F)}{4}$ & $\frac{\sqrt{3}(D+F)}{4}$ & $-\frac{\sqrt{3}(D+F)}{2\sqrt{2}}$ & $\frac{(D+2\,F)}{2}$ & $-\frac{D}{2}$\\

     $\bar K^*\Lambda$ & $\frac{(D-3\,F)}{4}$ & $\frac{1}{4}\frac{(D-3\,F)}{\sqrt{3}}$ & $-\frac{(D-3\,F)}{2\,\sqrt{6}}$ & $-\frac{D}{2}$ & $-\frac{D}{2}$\\

     \hline \hline
     
    \end{tabular}    
    \label{bij}
\end{table}

\begin{table}[h!]
    \centering
    \caption{$\mathcal{F}_{ij}$ coefficients used in Eq.~(\ref{krolleq}).}
    \begin{tabular}{c|p{2.6cm} p{2.6cm} p{2.6cm} p{2.6cm} p{2.6cm}}
    \hline \hline
     & $\rho\Xi$ & $\omega\Xi$ & $\phi\Xi$ & $\bar K^*\Sigma$ & $\bar K^*\Lambda$ \\ \hline

     $\pi\Xi$ & $2(F'-D')$ & $0$ & $0$ & $\frac{(D'+F')}{2}$ & $-\frac{(D'-3\,F')}{2}$\\

     $\eta\Xi$ & $0$ & $0$ & $0$ & $-\frac{3(D'+F')}{2}$ & $-\frac{(D'-3\,F')}{2}$ \\ 

     $\bar K\Sigma$ & $\frac{(D'+F')}{2}$ & $\frac{-\sqrt{3}(D'+F')}{2}$ & $\frac{\sqrt{3}(D'+F')}{2}$ & $-(D'+2\,F')$ & $D'$\\

     $\bar K \Lambda$ & $-\frac{(D'-3\,F')}{2}$ & $-\frac{(D'-3\,F')}{2
     \sqrt{3}}$ & $\frac{(D'-3\,F')}{\sqrt{6}}$ & $D'$ & $D'$\\

     \hline \hline
     
    \end{tabular}    
    \label{fij}
\end{table}

\begin{turnpage}
\begin{table}[h!]
    \centering
    \scriptsize
    \caption{$\mathcal{C}_{ij}$ coefficients used in Eq.~(\ref{VB:su}).}
    \begin{tabular}{c|c c c c c}
    \hline \hline
     & $\rho\Xi$ & $\omega\Xi$ & $\phi\Xi$ & $\bar K^*\Sigma$ & $\bar K^*\Lambda$ \\ \hline

     $\rho\Xi$ & $\frac{3(2\,\tilde{M} - (D-F)\,\tilde{m})^2}{16\,\tilde{M}^2}$ & $-\frac{\sqrt3(2\,\tilde{M} - (D-F)\,\tilde{m})^2}{16\,\tilde{M}^2}$ & $\sqrt\frac{3}{2}\Big(\frac{(D-F)F\,\tilde{m}^2}{4\tilde{M}^2} - \frac{F\,\tilde{m}}{2\tilde{M}} -1 \Big)+ \frac{3\tilde{m}(D-F)}{4\tilde{M}}$ & $\frac{3((F\,\tilde{m}+2\tilde{M})^2-D^2\tilde{m}^2)}{16\tilde{M}^2}$ & $-\frac{(D\,\tilde{m}-3F\,\tilde{m}-6\tilde{M})(D\,\tilde{m}-F\tilde{m}-2\tilde{M})}{16\tilde{M}^2}$ \\

     $\omega\Xi$ & $-\frac{\sqrt3(2\,\tilde{M} - (D-F)\,\tilde{m})^2}{16\,\tilde{M}^2}$ & $\frac{(2\,\tilde{M} - (D-F)\,\tilde{m})^2}{16\,\tilde{M}^2}$ & $-\frac{((D-F)\,\tilde{m}-2\tilde{M})(F\,\tilde{m}+2\tilde{M})}{4\sqrt{2}\tilde{M}^2}$ & -$\frac{\sqrt{3}((F\,\tilde{m}+2\tilde{M})^2-D^2\tilde{m}^2)}{16\tilde{M}^2}$ & $\frac{((D-3\,F)\tilde{m}-6\tilde{M})((D-F)\tilde{m}-2\tilde{M})}{16\sqrt{3}\tilde{M}^2}$\\ 

     $\phi\Xi$ & $\sqrt\frac{3}{2}\Big(\frac{(D-F)F\,\tilde{m}^2}{4\tilde{M}^2} - \frac{F\,\tilde{m}}{2\tilde{M}} -1 \Big)+ \frac{3\tilde{m}(D-F)}{4\tilde{M}}$ & $-\frac{((D-F)\,\tilde{m}-2\tilde{M})(F\,\tilde{m}+2\tilde{M})}{4\sqrt{2}\tilde{M}^2}$ & $\frac{(F\,\tilde{m}+2\tilde{M})^2}{2\tilde{M}^2}$ & $-\sqrt\frac{3}{2}\frac{(F\,\tilde{m}+2\tilde{M})((D+F)\tilde{m}+2\tilde{M})}{4\tilde{M}^2}$ & $-\frac{((D-3F)\tilde{m}-6\tilde{M})(F\,\tilde{m}+2\tilde{M})}{4\sqrt{6}\tilde{M}^2}$ \\

     $\bar K^*\Sigma$ & $\frac{3((F\,\tilde{m}+2\tilde{M})^2-D^2\tilde{m}^2)}{16\tilde{M}^2}$ & -$\frac{\sqrt{3}((F\,\tilde{m}+2\tilde{M})^2-D^2\tilde{m}^2)}{16\tilde{M}^2}$ & $-\sqrt\frac{3}{2}\frac{(F\,\tilde{m}+2\tilde{M})((D+F)\tilde{m}+2\tilde{M})}{4\tilde{M}^2}$ & $\frac{3((D+F)\tilde{m}+2\tilde{M}^2)}{16\tilde{M}^2}$ & $\frac{((D-3F)\tilde{m}-6\tilde{M})((D+F)\tilde{m}+2\tilde{M})}{16\tilde{M}^2}$\\

     $\bar K^*\Lambda$ & $-\frac{(D\,\tilde{m}-3F\,\tilde{m}-6\tilde{M})(D\,\tilde{m}-F\tilde{m}-2\tilde{M})}{16\tilde{M}^2}$ & $\frac{((D-3\,F)\tilde{m}-6\tilde{M})((D-F)\tilde{m}-2\tilde{M})}{16\sqrt{3}\tilde{M}^2}$ & $-\frac{((D-3F)\tilde{m}-6\tilde{M})(F\,\tilde{m}+2\tilde{M})}{4\sqrt{6}\tilde{M}^2}$ & $\frac{((D-3F)\tilde{m}-6\tilde{M})((D+F)\tilde{m}+2\tilde{M})}{16\tilde{M}^2}$ & $\frac{((D-3F)\tilde{m}-6\tilde{M})}{48\tilde{M}^2}$ \\

     \hline \hline
    \end{tabular}   
    \label{cij}
\end{table}
\end{turnpage}

\begin{turnpage}
\begin{table}[h!]
    \centering
    \scriptsize
    \caption{$\mathcal{D}_{ij}$ coefficients used in Eq.~(\ref{VB:su}).}
    \begin{tabular}{c|c c c c c}
    \hline \hline
     & $\rho\Xi$ & $\omega\Xi$ & $\phi\Xi$ & $\bar K^*\Sigma$ & $\bar K^*\Lambda$ \\ \hline
     $\rho\Xi$ & $-\frac{(2\tilde{M}-(D-F)\tilde{m})^2}{16\tilde{M}^2}$ & $-\frac{\sqrt 3 (2\tilde{M}-(D-F)\tilde{m})^2}{16\tilde{M}^2}$ & $\frac{\sqrt 3 ((D-F)\tilde{m}-2\tilde{M})(F\,\tilde{m}+2\tilde{M})}{4\sqrt 2 \tilde{M}^2}$ & $1+\frac{\tilde{m}((D^2+3\,DF+6F^2)\tilde{m}+6\tilde{M}(D+3F))}{24\tilde{M}^2}$ & $\frac{D\tilde{m}((D+F)\tilde{m}+2\tilde{M})}{8\tilde{M}^2}$ \\

     $\omega\Xi$ & $-\frac{\sqrt 3 (2\tilde{M}-(D-F)\tilde{m})^2}{16\tilde{M}^2}$& $\frac{(2\tilde{M}-(D-F)\tilde{m})^2}{16\tilde{M}^2}$ & $-\frac{((D-F)\tilde{m}-2\tilde{M})(F\,\tilde{m}+2\tilde{M})}{4\sqrt{2}\tilde{M}^2}$ & $-\frac{\sqrt 3\,((D-F)\tilde{m}+2\tilde{M})(F\,\tilde{m}+2\tilde{M})}{8 \tilde{M}^2}$ & $\frac{(D-3\,F)(2\,D-3\,F)\tilde{m}^2-6\tilde{m}\tilde{M}(3\,D-4\,F)+36\tilde{M}^2}{24 \sqrt 3 \tilde{M}^2}$ \\
     
     $\phi\Xi$ & $\frac{\sqrt 3 ((D-F)\tilde{m}-2\tilde{M})(F\,\tilde{m}+2\tilde{M})}{4\sqrt 2 \tilde{M}^2}$ & $-\frac{((D-F)\tilde{m}-2\tilde{M})(F\,\tilde{m}+2\tilde{M})}{4\sqrt{2}\tilde{M}^2}$ & $\frac{(F\,\tilde{m}+2\tilde{M})^2}{2\tilde{M}^2}$ & $\frac{\sqrt 3\,(D^2\tilde{m}^2-(F\,\tilde{m}+2\tilde{M})^2)}{8\sqrt 2 \tilde{M}^2}$ & $\frac{9\,(F\,\tilde{m}+2\tilde{M})^2-D^2\tilde{m}^2}{24 \sqrt 6 \tilde{M}^2}$ \\

     $\bar K^*\Sigma$ & $1+\frac{\tilde{m}((D^2+3\,DF+6F^2)\tilde{m}+6\tilde{M}(D+3F))}{24\tilde{M}^2}$ & $-\frac{\sqrt 3\,((D-F)\tilde{m}+2\tilde{M})(F\,\tilde{m}+2\tilde{M})}{8 \tilde{M}^2}$ & $\frac{\sqrt 3\,(D^2\tilde{m}^2-(F\,\tilde{m}+2\tilde{M})^2)}{8\sqrt 2 \tilde{M}^2}$ & $-\frac{(2\tilde{M}-(D-F)\tilde{m})^2}{16\tilde{M}^2}$ & $\frac{((D-F)\tilde{m}-2\tilde{M})((D+3\,F)\tilde{m}+6\tilde{M})}{16\tilde{M}^2}$ \\

     $\bar K^*\Lambda$ & $\frac{D\tilde{m}((D+F)\tilde{m}+2\tilde{M})}{8\tilde{M}^2}$ & $\frac{(D-3\,F)(2\,D-3\,F)\tilde{m}^2-6\tilde{m}\tilde{M}(3\,D-4\,F)+36\tilde{M}^2}{24 \sqrt 3 \tilde{M}^2}$ & $\frac{9\,(F\,\tilde{m}+2\tilde{M})^2-D^2\tilde{m}^2}{24 \sqrt 6 \tilde{M}^2}$ & $\frac{((D-F)\tilde{m}-2\tilde{M})((D+3\,F)\tilde{m}+6\tilde{M})}{16\tilde{M}^2}$ & $\frac{((D+3\,F)\tilde{m}+6\tilde{M})^2}{48 \tilde{M}^2}$

     \\

     \hline \hline
    \end{tabular}   
    \label{dij}
\end{table}
\end{turnpage}

\clearpage
\bibliographystyle{apsrev4-1}

\bibliography{RefsXi}

\end{document}